\documentclass[structabstract]{aa} 
\usepackage{graphicx}
\usepackage{txfonts}
\usepackage{lscape}
\usepackage{longtable}

\newcommand{\rot}{$v \sin i$ }

\newcommand{\ie}{{\em i.e.,} }
\newcommand{\eg}{{\em e.g.,} }
\newcommand{\cf}{{\em c.f.,} }
\newcommand{\caiihk}{Ca\,{\footnotesize II}~H \&~K }

\newcommand{\lii}{Li\,{\footnotesize I} }
\newcommand{\fei}{Fe\,{\footnotesize I} }

\newcommand{\lsi}{$\mathrel{\hbox{\rlap{\hbox{\lower2pt\hbox{$\sim$}}}\raise2pt\hbox{$<$}}}$}
\newcommand{\gsi}{$\mathrel{\hbox{\rlap{\hbox{\lower2pt\hbox{$\sim$}}}\raise2pt\hbox{$>$}}}$}
\begin{document}
   \title{New Rotation Periods in the Open Cluster 
NGC 1039 (M\,34), and a Derivation of its Gyrochronology 
Age.}

\titlerunning{The Gyrochronology Age of M\,34}


   \author{David J. James\inst{1,2}
          \and
           Sydney A. Barnes\inst{3}
           \and
           S{\o}ren Meibom\inst{4}
           \and
           G. Wesley Lockwood\inst{3}
           \and 
           Stephen E. Levine\inst{5}
           \and
           Constantine Deliyannis\inst{6}
           \and 
           Imants Platais\inst{7}
           \and
           Aaron Steinhauer\inst{8}  
           \and 
           Briana K. Hurley\inst{1}
        }

   \institute{H\={o}k\={u} Ke`a Observatory, Department of Physics \& Astronomy, 
University of Hawai`i at Hilo, 200 West Kawili Street, Hilo, HI 96720, 
USA: \email{david.james@hawaii.edu}
\and
Department of Physics \& Astronomy, Vanderbilt 
University, Box 1807 Station B, Nashville, TN 37235, USA
\and
Lowell Observatory, 1400 W. Mars Hill Rd., Flagstaff, 
AZ 86001, USA
\and 
Harvard-Smithsonian Center for Astrophysics, 
60 Garden Street, Cambridge, MA 02138, USA
\and 
United States Naval Observatory, Flagstaff Station, 
10391 West Naval Observatory Road, Flagstaff, AZ 86001-8521, USA
\and
Astronomy Department, Indiana University, Swain Hall 
West 319, 727 East 3rd Street, Bloomington, IN 47405-7105, USA
\and 
Department of Physics \& Astronomy, Johns-Hopkins University, Baltimore, 
MD 21218, USA
\and
Department of Physics and Astronomy, 1 College Circle, 
SUNY Geneseo, Geneseo, NY 14454, USA}

   \date{Received Aug 19, 2009; accepted Mar 16, 2010}

 
  \abstract
   {}
   {Employing photometric rotation periods 
    for solar-type stars in NGC 1039 [M\,34], 
    a young, nearby open cluster, we use its 
    mass-dependent rotation period distribution 
    to derive the cluster's age in a distance 
    independent way, i.e., the so-called gyrochronology 
    method.} 
   {We present an analysis of 55 new rotation periods, 
    using light curves derived from differential 
    photometry, for solar type stars in the open 
    cluster NGC 1039 [M\,34]. We also exploit the 
    results of a recently-completed, standardized, 
    homogeneous BVIc CCD survey of the cluster, 
    performed by the Indiana Group of the WIYN 
    open cluster survey, in order to establish photometric 
    cluster membership and assign B-V colours to each 
    photometric variable. 

    We describe a methodology for establishing the 
    gyrochronology age for an ensemble of solar-type 
    stars. Empirical relations between rotation period, 
    photometric colour and stellar age (gyrochronology) 
    are used to determine the age 
    of M\,34. Based on its position in a colour-period 
    diagram, each M\,34 member is designated as being 
    either a solid-body rotator ({\em interface or I-star}), 
    a differentially rotating star ({\em convective or 
    C-star}) or an object which is in some transitory 
    state in between the two ({\em gap or g-star}). 
    Fitting the period and photometric colour of each 
    I-sequence star in the cluster, we derive the 
    cluster's mean gyrochronology age.}
    {Of the photometric variable stars in the cluster field, for 
    which we derive a period, 47 out of 55 of them lie along 
    the loci of the cluster main sequence in V/B-V and V/V-I 
    space. We are further able to confirm kinematic membership 
    of the cluster for half of the periodic variables [21/55], 
    employing results from an on-going radial velocity survey 
    of the cluster. For each cluster member identified as an 
    I-sequence object in the colour-period diagram, we derive 
    its individual gyrochronology age, where the mean gyro 
    age of M\,34 is found to be $193 \pm  9$ Myr.}
    {Using differential photometry, members of a young open 
     cluster can be easily identified in a colour-magnitude 
     diagram from their periodic photometric variability alone. 
     Such periodicity can be used to establish a period-colour 
     distribution for the cluster, which for M\,34, we have 
     used to derive its gyrochronology age of $193 \pm  9$ Myr. 
     Formally, our gyro age of M\,34 is consistent (within the errors) 
     with that derived using several {\em distance-dependent}, photometric 
     isochrone methods ($250 \pm 67$ Myr). }


   \maketitle
%

\section{Introduction}\label{intro}

Over the past century, as instrumentation and detector technology 
have advanced, the methods by which stellar age is determined have 
also evolved. While we still rely heavily upon the use of theoretical 
model isochrones, fitting empirical luminosity-temperature data 
(on an H-R diagram - e.g., Sandage 1958; Demarque \& Larson 1964; 
Demarque \& Gisler 1975, Mengel et al. 1979; and more latterly, 
Meynet, Mermilliod \& Maeder 1993; D'Antona \& Mazzitelli 1997; 
Naylor 2009), several alternative techniques for determining stellar 
age now exist which exploit a more diverse assembly of fundamental 
stellar properties, such as magnetic activity, elemental abundances, 
white dwarf cooling time-scales and angular momentum content.

Inspired by the pioneering efforts of Wilson (1963), 
observations of solar type stars have shown that there 
is a clear age dependence in the strength of magnetic 
activity indicators (eg., \caiihk and X-rays), whose 
emissions are presumably due to solar-like magnetic 
field activity (eg., Wilson 1964; Wilson \& Skumanich 
1964; Wilson 1966; Skumanich 1972; Barry et al. 1981; 
Noyes et al. 1984; Maggio et al. 1987; Soderblom, 
Duncan \& Johnson 1991; Henry et al. 1996; Mamajek 
\& Hillenbrand 2008). While there is not a one-to-one 
relationship between activity and age for any given 
star, current data samples unequivocally show there 
to be a strong correlation between decreasing magnetic 
activity and increasing age.

However, even such a well-observed phenomenon does not 
deliver unequivocal results, as illustrated for instance 
by the factor of two range in ages derived by Giampapa et 
al. (2006) for photometric and kinematic members of the 
4Gyr-old open cluster M\,67. Moreover, stellar activity 
cycles, akin to those occurring in the Sun, can also lead 
to difficulties in establishing ages for individual stars. 
For instance, Soderblom et al. (1991) show that errors 
on chromospheric ages are roughly $50\%$. Presumably, this 
error could be driven down statistically in a cluster by 
measuring many stars, and by removing close binaries or 
other contaminants. Soderblom, Jones \& Fischer (2001) 
have provided chromospheric H$\alpha$ measurements for 
a significant number of M\,34 stars, but to our knowledge, 
a chromospheric age for M\,34 has not yet been published.

More recently, age determinations for young open 
clusters have been advanced based on an analysis of 
their light element abundance distributions, 
specifically using lithium as a tracer. This 
technique, the so-called lithium depletion boundary 
method (Basri, Marcy \& Graham 1996; Rebolo et al. 
1996; Stauffer, Schultz \& Kirkpatrick 1998; 
Barrado y Navascu\'{e}s, Stauffer \& Patten 1999), 
does however have both merits and limitations. 
Line strengths and elemental abundance ratios 
are fairly straight forward to measure, and are 
indeed fundamental stellar properties, perhaps 
more so than compared to magnetic activity indicators. 
Unfortunately, the strength of this technique lies 
in detecting lithium in M-stars and cooler, which for 
any group of stars at a given distance are among the 
faintest. Even for nearby open clusters, detecting 
the lithium boundary thus requires use of precious 
8-10m class telescope time. Perhaps the ultimate 
limitation for this technique is that it is only 
discernibly sensitive to stars of \lsi 250 Myr 
(\eg Stauffer et al. 1998), severely limiting its 
widespread applicability.

The use of heavy-element abundances to establish 
stellar age has also been attempted, however its 
implementation is as yet limited. Cayrel et al. 
(2001) advocate using the detection of the 
singly-ionized uranium 238 line at 3859.6\r{A} as 
an age proxy, although as they themselves note, 
the accuracy of this technique is currently limited 
by deficiencies in the nuclear data - \eg poorly-known 
oscillator strengths. Even with the advent of 
well-calibrated atomic data for uranium species, 
dating of open clusters using detection of stellar 
uranium will only be feasible for clusters older 
than the Hyades [600Myr], due to the prohibitively 
long half-lives of the two dominant naturally 
occurring uranium species ($\tau_{1/2}= 7.04\times10^{8}$ 
\& $4.47\times10^{9}$ yrs for U$^{235}$ \& U$^{238}$ 
respectively - CRC 2005).

Spectroscopic and photometric observations of white 
dwarfs, in the Galactic field and in globular and 
open clusters, can also yield an age estimate of 
the their stellar content. The detailed physics 
of how white dwarfs cool as they age is now well 
developed (Mestel 1952; Cox \& Giuli 1965; Beaudet, 
Petrosian \& Salpeter 1967), and their model-dependent 
cooling time-scales are well described using 
theoretical stellar models (Koester \& Sch\"{o}nberner 
1986; D'Antona \& Mazzitelli 1989; Iben \& Laughlin 1989; 
Wood 1992; Fontaine, Brassard \& Bergeron 2001). This 
technique has been historically rather difficult to 
implement in open clusters. First, even for the nearby 
clusters, their white dwarf members are the faintest 
objects comprising the mass function, typically being 
fainter than V=$20$th magnitude (\eg Rubin et al. 2008), 
which is challenging for all but the largest aperture 
terrestrial telescopes.  Second, white dwarfs are 
intrinsically blue objects (typically B-V$_{o}$~$<0.2$), 
which until recently was problematic for capturing 
photons efficiently using astronomical, red sensitive 
spectrographs and detectors. Third, and most 
importantly, to properly derive the age of the 
cluster one must observe and characterize objects 
extending right down to the bottom of the white 
dwarf sequence; \ie those objects which are the 
dimmest, the coolest and hence the {\em oldest} 
white dwarfs in the cluster. Consequently, in light 
of such procedural difficulties, relatively few 
open clusters have white dwarf cooling-time age 
estimates available in the literature (\eg Richer 
et al. 1998; Bedin et al. 2005; Kalirai et al. 2003, 
2005; Rubin et al. 2008).

In terms of rotation, Kawaler (1989) uses theoretical 
stellar spin-down models (from Kawaler 1988) to 
analytically derive a relationship uniquely connecting 
a star's B-V photometric colour and age to its period. 
Using a photometric colour, rotation period dataset 
for G \& K-stars in the Hyades open cluster, he shows 
the Hyades rotation age to be $4.9\pm1.1 \times 10^{8}$ 
years. More recently, Barnes (2003, 2007 - hereafter 
B03, B07 respectively) exploit the morphology of 
photometric colour-rotation period distributions in 
co\"{e}val samples of stars (\eg open clusters and 
binary systems) to establish stellar age. In essence, 
this so-called {\em gyrochronology} technique allows one to 
construct rotational isochrones, in order to trace the 
boundaries of age dependent colour-period distributions. 
Crucially however, these boundary definitions further 
suggest the identification and coupling-state of each 
star's basic internal stratification in terms of their 
radiative zone and convection envelope.

Solar-type stars lying close to or on the $I$, or 
interface sequence, are probably rotating as solid 
bodies or close to it, and their spin-down evolution 
is Skumanich-like (Skumanich 1972), i.e., directly 
proportional to the square root of stellar age 
(spin-down~$\propto ~t^{1/2}$). The most rapidly rotating 
solar-type stars at a given mass lie on or close to 
the $C$, or convective sequence. These stars are 
thought to be rotationally {\em de-coupled}; that 
is to say, it is likely that {\em only} their outer 
convective envelope is spinning down, with an 
exponential time dependency (spin-down~$\propto ~e^{f(t)}$). 
Early incarnations of rotational evolution models 
incorporating {\em de-coupled} stellar structure were 
referred to as core-envelope (de)coupling models (\cf  
Endal \& Sofia 1981; Stauffer et al. 1984; Soderblom 
et al. 1993a, Jianke \& Collier Cameron 1993). Those 
stars which are in a transitory state between the $C$ 
and $I$-sequences, representing a scenario whereby the 
outer convective envelope is {\em re-coupling}, probably 
magneto-hydrodynamically, to the inner radiative zone,  
constitute the so-called {\em gap} or $g-$stars in the 
gyrochronology paradigm.

Establishing stellar age for open clusters using 
gyrochronology models of course has its limitations. 
For instance for open clusters, enough stars must be 
photometrically monitored in order to derive rotation 
periods, so that a clear, well-defined distribution of 
interface, gap, or convective sequence objects is 
apparent. Such an observing programme typically requires 
extensive allocations of telescope time and considerable 
human effort. Moreover, differential rotation of solar-type 
stars and multiple star-spot groups on their surfaces 
can also introduce ambiguities into gyrochronology 
analyzes, since it can act to smear-out the distribution 
of rotation periods of a given mass. 

For small samples of rotation periods in open clusters, 
the gyrochronology method is not applicable in multiple 
systems where other effects might interfere with, or even 
overwhelm, the regular wind-related loss of angular momentum. 
For instance, close binaries ($a$ \lsi 0.1 AU) experience 
a tidal torque which acts to synchronize their rotation 
with the orbital motion of the system, in addition to any 
magnetic torque they experience due to stellar winds (Zahn 
1989; 1994); thus, as they evolve along the main-sequence, 
components in the closest binaries tend to rotate faster 
(on average) than single stars of similar mass and age. 
Observations of RS CVn and BY Dra systems have shown Zahn's 
theoretical framework to be correct (Hall 1976; Bopp \& 
Fekel 1977; Fekel, Moffet \& Henry 1986). Binary/multiple 
systems must therefore be avoided, where the data allow, 
in establishing gyrochronology ages of stars (see Meibom, 
Mathieu \& Stassun 2006 for a more thorough discussion).

A recent International Astronomical Union symposium (IAUS 258), 
dedicated to the subject of {\em The Age of Stars}, now provides 
our community with a thorough overview of an historical 
and modern approach to understanding the how and why of 
stellar ages. A careful perusal of the symposium proceedings 
allows, for both the novice and expert alike, a detailed 
insight into the theoretical framework unpinning stellar 
age determinations as well as the observational data 
calibrating and constraining such models. Of special 
interest to this manuscript were the oral presentations 
by Barnes (2009), Jeffries (2009) and Meibom (2009).

The genesis of this gyrochronology project for M\,34, a 
$\simeq$ 200\,Myr open cluster, occurred nearly ten years 
ago. At that epoch, SAB and GWL obtained a 
differential photometry dataset at Lowell Observatory  
over 17-nights for the Western half of the cluster, with 
the goal of deriving rotation periods for photometric 
variables in the field. The project lay dormant for some 
years until we noticed that Irwin et al. (2006) published 
new rotation period data for a sample of solar-type stars 
in the cluster. Unfortunately in terms of rotation, there 
are several fundamental problems with the Irwin et al. 
dataset which hamper our ability to pursue a rigorous 
gyrochronology analysis using their data (see 
\S~\ref{IRWINcomp} for details). A natural progression 
and indeed amelioration of their period distribution analysis 
is to derive the gyrochronology properties of this cluster, 
now incorporating rotation periods derived from the 
differential photometry that we had obtained during the 
1998 Lowell campaign. In this manuscript therefore, we now 
report an analysis of our differential photometry dataset 
for the cluster. 

New BVI photometry is reported in \S~\ref{indiana} for 
each photometric variable discovered in our Lowell observing 
campaign. Rotation periods derived from light-curve analysis 
of the photometric variables in our field of view are described 
in \S~\ref{newProt}, which includes a comparison with the 
Irwin et al. study. Using a clean sample of rotation periods 
for photometric and/or kinematic members of the M\,34 cluster, 
we describe in \S~\ref{period-distribution} how colour-period 
data are used in undertaking a gyrochronology analysis, which 
results in the derivation of the rotation age of the cluster. 
Also included is a commentary on how error in the gyrochronology 
age can be ascribed to disparate physical properties of the 
colour-period dataset, such as cluster non-membership, 
binarity and differential rotation. 

\section{The Open Cluster Messier\,34}\label{extantM34}

M\,34 (NGC\,1039) is a young ($\simeq$200 Myr; see 
Table~\ref{table-age}) open cluster on the border 
of Perseus and Andromeda [RA(2000): $02^{h}42^{m}$, 
DEC(2000): $+42^{o}46^{'}$]. It is $\simeq 470$ pc 
away, out of the Galactic plane [$b\simeq -16^{o}$], and 
has a reddening of E$_{B-V}$=0.07 (Canterna, Perry \& 
Crawford 1979). These characteristics suggested to us 
that it would make a good target for rotation period 
determinations, including a derivation of its rotational 
age via the methodology of gyrochronology.

\subsection{Extant Observations of M\,34}\label{extantOBS}

Prior work on M\,34 includes photoelectric $UBV$ 
observations for 57 stars bluer than F3 in the M\,34 
field by Johnson (1954) and of 43 similar stars by 
Cester et al. (1977). Canterna et al. (1979) reported 
photoelectric photometry in Str\"{o}mgren $uvby$ \& 
H$\beta$, for some 42 M\,34 stars. Jones \& Prosser (1996) 
published a more extensive and deeper (down to V=16.2) 
survey using $BVI$ CCD observations. An even deeper 
(V$\simeq 24$) $VI$ CCD survey has been recently 
published by Irwin et al. (2006). Both of the latter two 
studies have certain irregularities with the standardization 
procedures that are evident in the papers. Consequently, 
we use standardized photometry from a new $BVRI$ survey by 
the Indiana group of the WIYN open cluster study 
(Deliyannis et al. in prep) in our analysis 
(see \S~\ref{indiana}).

Two proper motion surveys have been conducted in M\,34. 
The first, by Ianna \& Schlemmer (1993), has a limiting 
magnitude of V=14.5, and the second, by Jones \& Prosser 
(1996), one of V=16.2. Reliable membership probabilities 
from these studies are limited to stars brighter than 
V=13.5 and V=14.5, respectively, and although $>$80\% of 
the stars in our study are fainter than either of the 
proper motion surveys, we use them where available. 
Jones et al. (1997) have provided radial velocities for 
some 47 cluster members but few of these overlap with our 
photometric period sample. However, we take recourse 
to an on-going multi-epoch WIYN HYDRA radial velocity 
survey of the cluster (Meibom et al. in prep) to provide 
a reliable kinematic membership criterion for nearly 
half (25/55) of the M\,34 stars for which we report 
new photometric periods. A thorough discussion of extant 
membership probabilities for M\,34 stars comprising 
our gyrochronology sample, as well as new high 
resolution spectroscopic results for two photometric 
variables in our sample, is presented in 
Appendix~\ref{J97mem}.

Other meritorious work on M\,34 includes a study of the 
cluster white dwarfs by Rubin et al. (2008), who derive a 
minimum age of the cluster to be 64~Myr. Although Rubin et 
al. were not explicitly attempting an age derivation, such 
a result must be considered with some caution because of 
the lack of surety of cluster membership for their targets, 
the small sample size (5 objects), and the fact that only 
the very top of the cluster white dwarf sequence was sampled, 
whereby the coolest, and hence oldest white dwarfs in the 
cluster were not observed.

Projected equatorial rotation velocities [v $\sin i$] 
for the earlier-type, higher mass stars in M\,34 are 
provided by Ianna (1970), Hartoog (1977), and Landstreet 
et al. (2008), whereas for its solar-type members 
velocities are published by Jones et al. (1997) and 
Soderblom et al. (2001). For stars in common, we provide 
a comparison between these $v \sin i$ data and our new 
photometric periods in Appendix~\ref{Vsini-comp-period}. 
By removing the dual ambiguities in $v \sin i$ data (the 
obvious $\sin i$ factor, and the stellar radius), photometric 
rotation period studies far and above supersede $v \sin i$ 
studies - provided of course that the rotation periods are 
measured correctly.

Irwin et al. (2006) have provided rotation periods 
for some 105 stars in M\,34, using photometric 
data obtained under considerably more restrictive 
conditions (observing window, and filters) than ours. 
Although half the stars in common between their study 
and our Lowell sample yield essentially the same period 
from multiple time-series, some discrepancies exist. 
A detailed comparison between their periods and ours 
is presented in \S~\ref{IRWINcomp}.

\section{New Indiana CCD photometry}\label{indiana}

A new, comprehensive BVRI survey of star fields 
in the direction of the M\,34 cluster is now available 
from the Indiana Group of the WIYN Open Cluster 
Survey (Deliyannis et al. in prep). Photometric observations 
of the central $40^{'}\times40^{'}$ of the cluster, 
including $\simeq 100$ Landolt (1992) standard stars, 
were obtained using the WIYN 0.9-meter telescope 
during the night of 24-Oct-1998. The $2048\times2048$ 
pixel T2KA Tekronics CCD was employed at 
the f/7.5 Cassegrain focus, which with its $0~\farcs68$ 
pixel$^{-1}$ plate scale yields an effective 
field of view of $\simeq 23^{'}\times23^{'}$. Multiple 
exposures in each filter were obtained over a 2x2 
mosaic for the cluster. Images were analyzed using 
the dedicated DAOPHOT II photometry data reduction 
package. Landolt standard star frames were processed 
using aperture photometry, whereas object frames were 
fitted with spatially-variable point spread functions 
[PSFs], and spatially-variable aperture corrections. 
Both the PSFs and aperture corrections were 
determined empirically using hundreds of isolated stars 
on each frame. External errors in placing instrumental 
magnitudes onto the standard system are $< 0.02$ magnitudes. 

The results of the Indiana Group's photometric 
survey for stars relevant to this study are 
presented in Table~\ref{table-identifier}, as well as 
coordinates for each star derived from SuperCosmos 
survey images (Hambly et al. 2001a, b; Hambly, 
Irwin \& MacGillivray 2001). We also provide a 
kinematic membership assessment from an on-going 
survey of the cluster (Meibom et al. in prep). 
For two M\,34 targets for which we derive a period, 
F4$\_1925$ \& F4$\_2226$, there are no $B$-band photometric 
data available. In this instance, we rely upon a field 
star V$-$I to B$-$V relationship (Caldwell et al. 1993), 
and an assumed E$_{B-V}$=0.07 (Canterna et al. 1979), 
to transform these stars' V-Ic colours into B-V ones. 

\section{New Photometric Periods in M\,34}\label{newProt}

\subsection{Observational Programme}

We initiated a differential photometry programme 
of observations for solar-type stars in M\,34 
during September-October 1998, consisting of 
four separate sequences of CCD images in 
the V and I filters, acquired respectively during 
a 17 night observing run (UT980929 through 981015) 
at Lowell Observatory, Flagstaff, Arizona, USA. The 
dataset consists of $\sim 60$ visitations to each of 
two 19'$\times$19' fields of the Western region of 
M\,34. Each visitation typically consisted of short 
({\bf 60, 40 sec}) and long ({\bf 600, 400 sec}) 
exposures in the V and I filters respectively, 
using the 2048$\times$2048 SITe chip mounted 
at the Cassegrain focus of the 42inch Hall 
telescope of Lowell Observatory. 
Each field received 4-7 visitations 
per night in each filter, over the first 12 nights, 
barring UT981008 which was lost to clouds, and one 
visitation each on the remaining 5 nights. The 
time-series analysis is performed on the instrumental 
system with respect to chosen reference images on 
UT981001. 

\subsection{Data Analysis and Time Series Analysis:}

The individual images were de-biased and flat-fielded 
using standard procedures in the IRAF\footnote{IRAF 
is distributed by the National Optical Astronomy 
Observatories, which are operated by the Association 
of Universities for Research in Astronomy, Inc., 
under cooperative agreement with the National Science 
Foundation.} image reduction utility. Photometry for 
all objects on the processed images was computed via 
PSF fitting using DAOPHOT II and ALLSTAR II 
(Stetson, Davis \& Crabtree 1991). 
Typically, we have used 50 PSF stars per image 
and a quadratically-varying PSF. The photometry 
from the individual frames was cross-matched and converted 
into time series using the DAOMATCH/DAOMASTER 
routines from Stetson (1992). These routines match 
stars appearing in multiple images according to 
user-supplied specifications, calculate the photometric 
offsets between frames, and output the corrected time 
series. The uncertainty in the frame-to-frame magnitude 
offsets was less than 0.001 mag for the frames that 
were eventually used because of the large number 
(\gsi 1000) of stars matched from frame to frame. 
This procedure left us with totals of $\sim 4000$, 
$\sim 3000$, $\sim 2700$, and $\sim 2200$ objects 
respectively with time series in each of the long-I, 
long-V, short-I and short-V exposures for the western 
M\,34 fields.

We tested every object in the field for variability 
without regard to its position in a colour-magnitude 
diagram, or any other consideration except its 
intrinsic variability. Also, we performed the 
variability/periodicity analysis independently in 
each exposure time/filter combination, in order to 
confirm or deny it, and to calculate period error. 
The following procedure was adopted for each 
time-series file. The time series files for each 
field were sigma-clipped to eliminate outliers\footnote{This 
procedure lowers our sensitivity to eclipsing binaries 
and flares or other one-time variables, but has the 
virtue of eliminating occasional bad photometric 
points, which could reduce the efficacy of the 
period-searching software for rotational variables.}, 
and then we calculated a reduced chi square 
($\tilde{\chi}^2$) for each time-series, to glean out 
the most variable objects. Every object with 
$\tilde{\chi}^2 > 3$ was subjected to a periodicity search 
using the CLEAN algorithm of Roberts, Lehar 
\& Dreher (1987). The results were examined visually, 
and phased about reliable peaks in the power spectrum 
to yield phased light curves for the periodic 
variables. Finally, results of the (up to) four 
independent time-series analyzes on each object 
were merged together to confirm or deny the 
periodicity. For these stars, we calculate period 
errors based on multiple measurements.

As a result of the foregoing analysis, we have been 
able to identify forty-seven (47) objects with 
periods confirmed in at least two filter, exposure 
combinations. In addition, we derive periods for 
eight (8) objects for which we can only detect 
periodicity in one filter/exposure. The period 
analysis results are listed in Table~\ref{table-Prot}, 
which also contains details of their derived periods, 
and standard deviations of magnitude variation in 
each time-series. Stars having only one period 
determination are assigned period errors based 
on the widths of the peaks in the power spectra. 
Phased photometric light-curves, on the instrumental 
system, are presented for each of our M\,34 variables 
in Appendix~\ref{lightcurves}.

Our 17-night observing window allows us to detect two 
or more phases for all periods up to about 8.5~days. 
However this cut is probably too conservative, and because 
the slow rotators are of great interest to us, we extend 
our analysis to periods longer than 8.5~days. We are 
confident about periods as long as three quarters of 
the 17d baseline, i.e. $\sim 13$d; however, the period 
errors are larger for such stars. For instance, F3\_0469 
(field 3) has a period of $\sim 12.9$d in all 4 exposure 
time/filter time-series. Periods longer than our 13-day 
{\em comfort} timescale abut the sensitivity limit of 
our observing window, which result in consequentially 
larger period errors for the [2/55] systems for which 
we report periods $>$13 days (see Table~\ref{table-Prot}).

\subsection{Establishing Membership of M\,34 from Photometric 
Variability Alone}\label{Protmembership}

Constructing a colour-magnitude diagram is the simplest 
way of establishing possible membership of an open cluster 
because stars located far from the cluster sequence are 
guaranteed to be non-members. Stars on the cluster 
sequence may or may not be {\em bona fide} cluster 
members, depending on the extent of fore- and background 
contamination from the Galactic field. For stars forming 
an apparent cluster main sequence, additional information 
is needed to distinguish between true members and field-star, 
non-member interlopers. Classical, well established methods 
for member, non-member triage involve astrometric (proper motion) 
measurements or measuring radial velocities (\eg Stauffer 1984, 
Meynet et al. 1993, Perryman et al. 1998, Raboud \& 
Mermilliod 1998, Mermilliod et al. 2008), however such 
measurements are not always practical to acquire. In 
their absence, magnetic activity (\eg X-rays) or lithium 
abundance indicators can be used to distinguish between 
cluster and non-cluster stars (\eg Prosser et al. 1995a; 
James \& Jeffries 1997; Cargile, James \& Platais 2009). 
Here, we propose that photometric periodicity can also be 
exploited to easily establish open cluster membership in a 
manner unbiased with respect to brightness, mass (at least 
for solar-type stars) and distance.

In Fig.,~\ref{CMDs}, we plot colour magnitude diagrams of 
the cluster in BV and VI colours, along with D'Antona \& 
Mazzitelli (1997) theoretical isochrones for 250\,Myr \& 470\,pc 
in both colours. The 55 periodic variables in our Lowell fields 
are also identified on the figure. Photometric and/or radial 
velocity members, 43 in number, are flagged with red squares, 
whereas non-members, 12 in number, are flagged with blue circles 
(their periodicity and other information are tabulated in 
Tables~\ref{table-identifier}~\&~\ref{table-Prot}). The cluster 
sequence is traced out very well by the periodic variables. In fact, 
in this particular case, photometric periodicity information alone 
provides a 78\% ($43/55$) probability of cluster membership. We 
believe the reason for this behavior is that the cluster stars 
are younger and hence far more active and variable than the mostly 
older disk stars in the field. One might even envision using 
photometric variability to trace out cluster sequences in sparsely 
populated young clusters or background-dominated cluster fields.

\subsection{Comparison with prior periods}\label{IRWINcomp}

%
%

%
%
%
%

Irwin et al. (2006) have published a prior study of periodic 
variables in M\,34, and before proceeding further, we compare 
our periods with theirs for stars in common between the two 
studies. This comparison is graphically represented in 
Fig.,~\ref{IRWINcompFIG}, where symbols are assigned to stars 
having a varying number of independent period derivations (1-4) 
in our dataset. Of the 19 stars in common between the Irwin et al. 
sample and our own Lowell campaign, roughly half (9/19) have 
periods that agree within 10\%. For all but two of the remaining 
10 stars, Irwin et al. periods are significantly shorter than 
ours, where the origins for such discrepancies are not entirely 
clear to us. 

Through comparison with an as yet unpublished photometric 
variability study of M\,34 we can establish confidence in the 
Lowell campaign periods, and advocate their preferential use 
compared to the Irwin et al. ones. Three stars in common to the 
Irwin et al. study and our Lowell campaign, F3\_0413, F4\_0234, 
and F4\_1147, have Meibom et al. (in prep) periods, whose 
photometric variability was characterized over several months 
worth of observation data (\eg see Meibom et al. 2009 for 
details of a similar observing programme for the M\,35 cluster). 
These data, flagged stars in Fig.,~\ref{IRWINcompFIG} and 
listed in Table~\ref{soren-syd-irwin-3stars}, show that the 
Meibom et al. periods are an excellent match to the Lowell 
periods (to within 1\%), whereas the Irwin et al. periods are 
poorly matched.

Proving the correctness of one period derivation over 
another is not an easy task. However, we note that our dataset 
has a 17-night baseline compared with the 10-night baseline of 
Irwin et al., suggesting that we have better sensitivity to 
longer period systems. Our data were also acquired all night, 
rather than restricted to half nights as the Irwin et al. 
observational data were. These facts strongly suggest that 
our data are less sensitive to period aliasing; indeed, 
several of the Irwin et al. stars lie on or close to particular 
alias curves, as shown in Fig.,~\ref{IRWINcompFIG}. Finally, 
we note that most of our periods [17/19] are confirmed in 
at least a second filter/exposure time combination in the 
Lowell campaign data. Consequently, in light of the concerns 
that we have for aliasing effects in some of the Irwin et al. 
data, we will preferentially use our own periods for the 
remainder of this work.

\section{The Photometric Period Distribution in M\,34}\label{period-distribution}


Our rotation period distribution for photometric and/or kinematic 
members of M\,34, as a function of intrinsic $B-V$ and $V-I$ colours, 
is displayed in Fig.,~\ref{M34-Prot-BVoVIo}. In agreement with prior 
results in various open clusters (B03, B07), there are no stars located 
in the upper left portion of the colour-period diagrams. The Sun-like 
[(B-V)$_{o}\simeq 0.64$] stars in our M\,34 sample are rotating about 
an order of magnitude faster than the Sun. However the lower mass stars, 
mid-K and later, have an order-of-magnitude spread in rotation period, 
as has been seen before in other young open clusters. Further detailed 
interpretation of the M\,34 colour-period diagram requires a comparison 
of the colour-period distributions for other young clusters.

Rotation period distributions for two comparison 
clusters whose ages straddle that of M\,34 will 
allow us to critically investigate the 
gyrochronological properties of M\,34. To this 
end, we display colour-period diagrams for the 
$\simeq$135 Myr M\,35 and 600~Myr Hyades clusters 
in Fig.,~\ref{CPD-M35-Hyades}. Colour-period data 
are taken from the recent study of M\,35 by 
Meibom et al. (2009), and for the Hyades, from 
Radick et al. (1987) and Prosser et al. (1995b). 
This side-by-side comparison shows that the 
$\sim$135\,Myr-old M\,35 cluster has two distinct 
sequences of rotating stars indicated by the 
solid and dashed line in the figure, while the 
older Hyades cluster has only one. B03 called 
these sequences `I' (for Interface) and `C' 
(for Convective), with `g' (for gap) stars in 
transition from C to I (see \S~\ref{intro} for 
a discussion of the angular momentum characteristics 
of CgI stars). What we see plainly from the 
comparison is that by the Hyades age, most 
rotating solar-type stars have been transformed 
into I-type stars. Graphs of this C-g-I transformation 
for a series of clusters can be found in B03 and 
Meibom et al. (2009).

We now take these gyrochronology sequences from the 
M\,35 colour-period distribution, and overplot them 
on the M\,34 data in the left hand panel of 
Fig.,~\ref{M35-ICseq-M34-CPD}. This allows us to 
classify the gyrochronology status of the (far fewer) 
M\,34 stars roughly, as indicated by the CgI symbols 
in the figure. We note that its C\,sequence is in 
the same approximate position as that of M\,35, but 
its I\,sequence lies above that of M35, as expected 
from its older age. The manner of this inter-comparison 
suggests that only the stars shown in the right panel 
of the figure ought to be identified as I-sequence 
stars in M\,34, which we henceforth do.

\subsection{A Gyrochronology Age for M\,34}\label{M34gyro}

It is desirable to derive an age for M\,34 using our 
photometric period dataset, and the method of 
gyrochronology (B03, B07) will allow us to pursue such 
an analysis. This exercise will allow us to examine 
whether the gyro age derived is physically reasonable, 
and hence whether it might be usable for other clusters 
in future. The basis for such a gyro age calculation 
relies upon; 

\begin{enumerate}
\item Deriving the true rotation period distribution of the cluster,
\item Removing possible cluster non-members, and
\item Choosing, and using only, the I-sequence stars in the cluster for the analysis.
\end{enumerate}

\begin{figure*}
\includegraphics[angle=-90,scale=0.65]{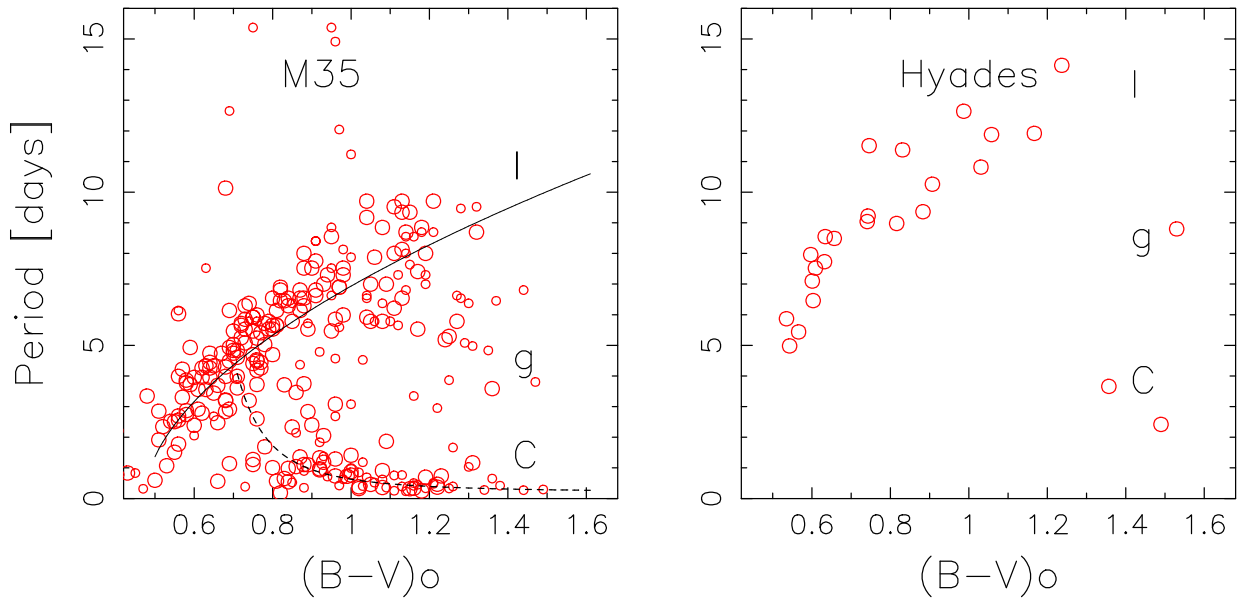}
\caption{In the left-hand panel, the rotation period distribution 
(in days) of the $\simeq$135 Myr M\,35 cluster is plotted 
against intrinsic (B-V) colour (data from Meibom et al. 2009); 
stars which are both photometric and kinematic members of 
the cluster are plotted with larger symbols than stars having 
photometric membership alone. Over-plotted are 135~Myr interface 
[solid line] and convection [dashed line] gyrochronology loci 
from the calibration by Meibom et al. (2009). In the right-hand 
panel, similar colour-period data are plotted for the 600 Myr 
Hyades cluster (data from Radick et al. 1987; Prosser et al. 1995b). 
Under the paradigm of gyrochronology, specific regions of the 
colour-period distributions are identified with C [convective], g 
[gap] and I [interface] symbols.}
\label{CPD-M35-Hyades}
\end{figure*}

\begin{figure*}
\includegraphics[angle=-90,scale=0.65]{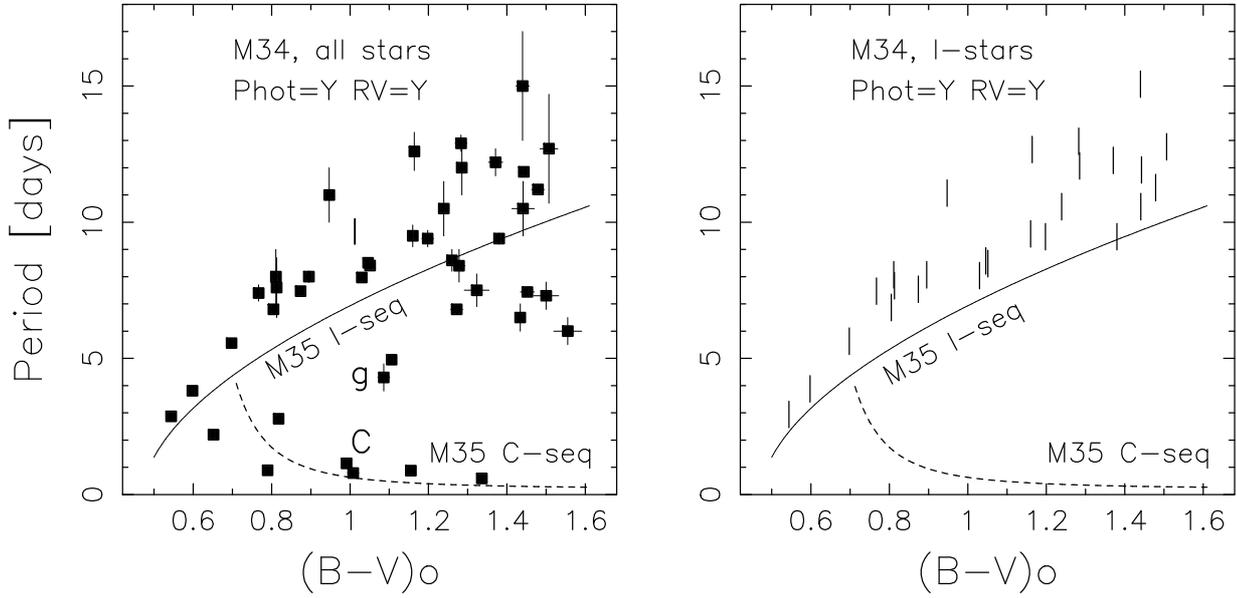}
\caption{The colour-period distribution for all photometric and/or 
kinematic members of M\,34 (see Table~\ref{table-Prot}) is plotted 
in the left hand panel. Over-plotted are gyrochronology Interface and 
Convection sequence loci (solid \& dashed lines respectively) 
for the 135~Myr M\,35 cluster (Barnes 2003, Meibom et al. 2009). 
Regions of the colour-period diagram where approximately 
200~Myr stars, classified as Interface, gap and Convective 
status in gyrochronology space, are identified with I, g or C 
symbols respectively. Candidate Interface sequence stars in 
the M\,34 cluster, lying above the Meibom et al. 
M\,35 Interface sequence, are identified as I-symbols 
in the right-hand panel; see \S~\ref{period-distribution} 
for details.}
\label{M35-ICseq-M34-CPD}
\end{figure*}

For M\,34, we have performed these three steps to the best of 
our ability, as described above, with the resulting set of candidate 
I-sequence stars in the cluster being displayed in the right-hand 
panel of Fig.,~\ref{M35-ICseq-M34-CPD}. We note that the 
forthcoming study by Meibom et al. (in prep), which contains 
many more M\,34 stars, will make the CgI distinction in this 
cluster more obvious, and verify our classifications made here. 
Based on fitting the open cluster colour-period distributions 
then available, B07 suggested that the positions of the 
I\,sequences of open clusters, or indeed that of field stars, 
follows the relationship;

\begin{equation}
\mathrm{P}_{rot} = f(B-V).g(t)
\end{equation}

where P$_{rot}$, (B-V), and $t$ represent the measured rotation 
period, measured de-reddened colour, and the age respectively, 
and $f$ and $g$ are fitted functions of the stellar colour 
and age respectively, where he proposed that

\begin{equation}
f(B-V) = a (B-V-c)^b 
\end{equation}

and

\begin{equation}
g(t) = t^n
\end{equation}

for which Barnes (2007) derived coefficients of $a=0.7725 \pm 0.0110$, 
$b=0.601 \pm 0.024$, $c=0.4$, and $n=0.5189 \pm 0.0070$. 
However, based on the most extensive open cluster rotation 
period study to date, that of M\,35 by Meibom et al. (2009 - 
displayed in the left-hand panel of Fig.,~\ref{CPD-M35-Hyades}), 
the coefficients of the I-sequence fit have been recently 
updated to be $a=0.770 \pm 0.014$, $b=0.553 \pm 0.052$, and 
$c=0.472\pm0.027$ with a fixed $n=0.52$. However, the index 
$n$ is set by the solar calibration, by demanding that the 
gyrochronology relationship for the I-sequence yields the 
solar rotation period at the solar age, the rationale for 
which is explained at length in B07. We note that, using 
their extensive rotation period dataset for M\,35, Meibom 
et al. re-fit the colour function coefficients of Equ.,~2 
but did not re-fit the age exponent, $n$, of Equ.,~3. 
Therefore to be mathematically rigorous, we employ the 
Meibom et al. (2009) $a, b$ \& $c$ coefficients and 
the solar data summarized in Barnes (2007) to re-derive 
the $g$(t) exponent to be $n=0.5344\pm0.0015$.

Taking the Meibom et al. (2009) $a, b$ \& $c$ coefficients 
and the newly derived $n$ exponent, we fit the Equ.,~1 
relationship to the colour-period data of M\,34 I-sequence 
stars, those identified in Fig.,~\ref{M35-ICseq-M34-CPD}, 
in order to determine the cluster's gyro age, the one free 
parameter in the fit. The results of this procedure are 
reported in Table~\ref{Istars-table-Prot}. An unweighted 
fit to these data yields a rotation age for M\,34 to be 
$193 \pm  9$ Myr\footnote{Using all of the Meibom et al. 
(2009) coefficients, including $n=0.52$, a resultant 
gyrochronology age for M\,34 of $223 \pm 11$ Myr would be 
obtained.} ($\pm$ possible systematic errors in gyrochronology), 
where the quoted error is the error on the mean for these 
data. Such a gyro age for the cluster is consistent with 
the Jones \& Prosser and Meynet et al. isochronal ages for 
M\,34.

\subsection{Variance of the Derived Periods about the Fit}\label{GYRO-errors}

Let us see if we can clarify the origin of items in the 
error budget contributing to the uncertainty in the cluster's 
gyrochronology age, which are in addition to any systematic 
error of the gyrochronology method itself. We can estimate 
the magnitude of the error budget by considering the additive 
contribution of each star's period variance\footnote{We define 
each star's period variance as (Measured Period - Fitted 
Period)$^{2}$, where the measured period is the photometric 
variability period determined from our differential photometry 
and the fitted period is that derived from Equ~1, using 
dereddened B-V colours, an assumed mean gyro age of 193~Myr 
and the Meibom et al. (2009) $a, b, c$ coefficients, with 
an $n$ exponent, $n=0.5344\pm0.0015$.}. We might expect 
contributions to the period variance from cluster non-members, 
tidally-interacting binaries within the cluster, a possible 
age spread within the cluster, period errors, differential 
rotation, and finally, one from initial variations in the 
cluster's natal angular momentum distribution. The data 
listed in Table~\ref{Istars-table-Prot} show that the total 
period variance about the fit is $\sim 43.7$ days${^2}$.


It is beyond the scope of this paper to consider possible 
systematic errors in the gyrochronology method, which is 
calibrated using both the Sun and a selection of young open 
clusters with measured photometric periods. An understanding 
of this error will probably require another decade of 
rotational work on open clusters. It would be unsurprising 
if such a study eventually identified a 15-20\% systematic 
error in the application of gyrochronology to real colour-period 
samples. 

The cluster colour-magnitude diagrams (see Fig.\,~\ref{CMDs}) 
suggest that 3 periodic field variables could have sneaked 
into our 43 member star sample had we not had the radial velocity 
measurements to reject them. This suggests that of our 22 
purely photometrically chosen members (see Table~\ref{table-Prot}), 
one possibly two are non-members, making this unlikely to be 
any significant contributor to the error. The mean variance 
contributor of any given star in our M\,34 period sample is 
1.68 days$^{2}$ (see Table~\ref{Istars-table-Prot}), which 
means that non-member variance contribution from photometric 
members is at most $\simeq 3.4$ days$^{2}$ ($\sim 8\%$ of the 
variance total). Of the 26 candidate I-sequence stars in M\,34, 
twelve of them are classified as photometric-only members of the 
cluster, \ie having no kinematic classification (with a mean 
period variance of 2.33 days$^{2}$). Of these twelve, only three 
have noticeably large period variances (F4\_1357, 7.84 days$^{2}$; 
F4\_1007, 5.76 days$^{2}$; F3\_0664, 4.41 days$^{2}$). In any case 
contribution to period variance, and hence gyro age errors, by 
cluster non-members is at a low level.



There is also a possibility that some scatter in the gyro ages 
for individual M\,34 stars is contributed by systems in tidally 
locked binaries, which will be addressed in greater detail in 
the Meibom et al. (in prep) study. We note that of the candidate 
I-sequence stars reported in Table~\ref{Istars-table-Prot}, 14/24 
have kinematic classifications. Only one of these stars is a binary 
system, with no data being available as to whether it is 
tidally-locked or not. On average, one such data point would 
contribute 1.68 days$^{2}$ to the total variance, or $\sim 4\%$.

Observational studies of other young open clusters and associations 
suggest that they are c\"{o}eval to within about a few Myr 
(\eg Wichmann et al. 2000; Palla \& Randich 2005; James et al. 2006). 
While spreads in observed quantities that also change with age 
are generally believed to originate in other ways, we note that 
recent theoretical work by Baraffe, Chabrier \& Gallardo (2009) 
shows that as a phenomenon, episodic accretion onto young 
solar-type stars can also act to introduce a luminosity 
spread, and hence age spread, in the main sequence loci of 
Hertzsprung Russell diagrams. An intrinsic age spread of 
10Myr (\eg Jeffries 2009) at B-Vo colours of 0.60, 1.00 \& 
1.40 in our 193Myr gyrochronology fit yields a maximum 
period variance of only 0.1 days$^{2}$. Such an intrinsic 
internal age spread within the cluster will therefore not 
contribute any significant variance.



\begin{figure*}
\includegraphics[angle=-90,scale=0.65]{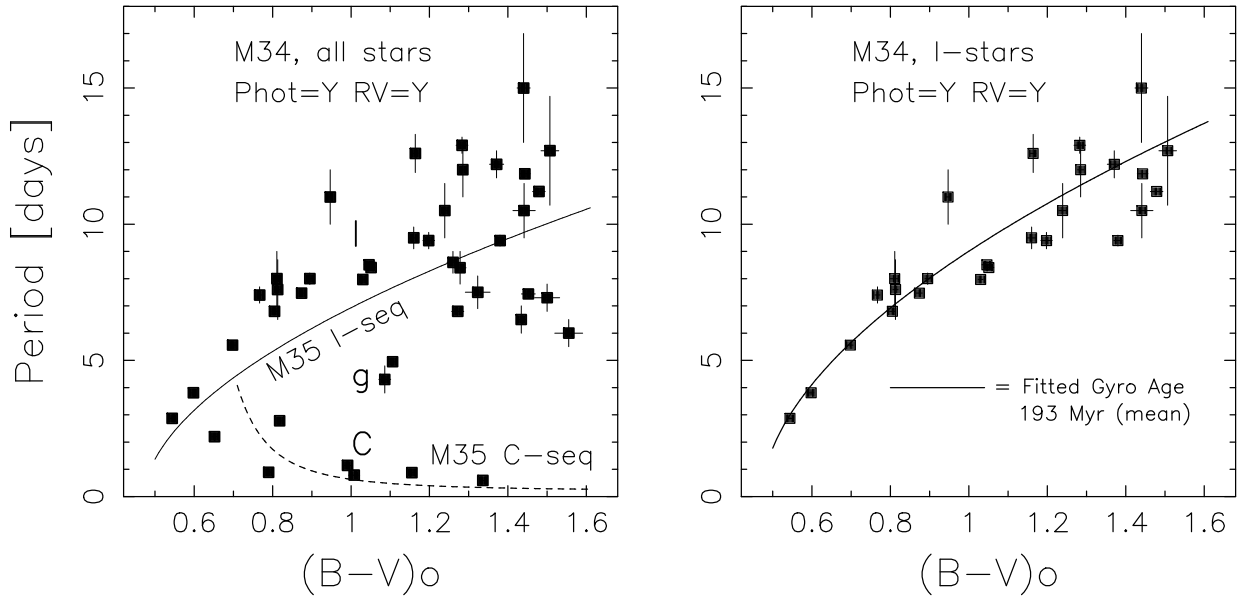}
\caption{This figure, very similar to Fig.,~\ref{M35-ICseq-M34-CPD}, shows 
the colour-period data for our M\,34 candidate I-sequence stars. The 
additional solid line in the right-hand panel represents the mean 
gyrochronology age of the I-stars in M\,34 (193 Myr - see 
Table~\ref{Istars-table-Prot}). This new gyrochronology I-sequence 
is calculated using updated colour-function coefficients (see Equ.,1-3 
and Meibom et al. 2009) together with our new derivation of the $g$(t) 
exponent ($n=0.5344$ - see Equ.,~3). }
\label{M34-Iseq-193Myr}
\end{figure*}

Inspection of Fig.\,\ref{M35-ICseq-M34-CPD} and 
Table~\ref{table-Prot} shows that photometric period 
errors are quite large for the longer periods in our 
sample. We attribute this effect to the 17-night baseline 
of observations and expect that these errors will decline 
with a longer baseline dataset, such as the forthcoming 
Meibom et al. (in prep) sample. Indeed, an informal comparison 
shows that they will be able to define the cluster's I-sequence 
better than our dataset allows. Another indication of 
this is that M\,34 I-sequence stars with smaller periods 
errors are, in general, located closer to the fitted 
sequence (see Fig.~\ref{M34-Iseq-193Myr}). Based solely 
on the period measurement errors themselves (see 
Table~\ref{Istars-table-Prot}), the variance contribution 
due to imprecise period measurement is $\simeq$ 15.5 days$^{2}$ 
(35.5\% of the total).


Differential rotation must be present in this sample, in 
the sense that two stars with the same equatorial velocity 
might present spot groups at observed at different latitudes, 
introducing some period scatter in the observations. We do 
not however expect this to contribute much to the observed 
variance, based on the parametrization of its contribution 
to the fractional period variance via 
$\delta P/P = 10^{-1.85} P^{0.3}$, following the analysis 
in Donahue, Saar \& Baliunas (1996) and Barnes (2007). For 
the I-sequence stars listed in Table~\ref{Istars-table-Prot}, 
this effect contributes only something like 
$\simeq 2.06$ days${^2}$ to the variance, equivalent to 
$5\%$ of the total variance budget.

This leaves a remainder variance of $21.1$ days${^2}$ 
(48\% of the total) to be accounted for by the effects 
of either; (a) the systematic difference between the 
gyro fit and the observation data, or (b) the residual 
influence of initial variations. Because these two 
contributors have differing dependencies on cluster 
and observational parameters, it should be possible 
to understand their separate contributions as the 
rotation period database of open cluster stars expands.

\section{Discussion and Summary}\label{discussion}

We have presented the results of a 17-night 
differential photometry campaign over the 
Western half of the central region of the 
open cluster M\,34. For all photometrically 
variable stars, we construct differential 
photometry light-curves, from which we derive 
periodicity. It is assumed that the photometric 
variability of objects consistent with cluster 
membership is due to the presence of magnetic-field 
induced starspots on their surfaces, rotating 
with the star; the derived periods are therefore 
representative of the angular velocity of the star.

In order to assess cluster membership, we exploit 
an extensive standardized CCD survey of 
the cluster, which shows that the majority of 
photometric variable stars [47/55] lie along the 
loci of the cluster main sequence in V/B-V and 
V/V-I space. Moreover, we are able to confirm 
kinematic membership of the cluster for 21 stars 
from an on-going radial velocity survey of the 
cluster (Meibom et al. in prep), 5 of which show 
radial velocity variations (multiple systems). 
Four (4) of the photometric variables are kinematic 
non-members. Stars which are either photometric 
or kinematic non-members were excluded from the 
gyrochronology analysis.

We note the existence of another photometric period 
dataset for solar-type stars in the M\,34 cluster 
(Irwin et al. 2006). We present an analysis 
comparing periods for stars in common between our 
Lowell campaign and their study. We detail several 
concerns that we have with the integrity of their 
dataset, notably their short observing window 
($<10$ nights) and their half-night observing 
cadence. After having verified that several 
Irwin et al. periods are most likely measurement 
aliases, and that our Lowell campaign periods 
agree those derived from an independent, separate, 
long time base-line differential photometry survey, 
we preferentially employ our period results in 
pursuing a gyrochronology age analysis for this 
cluster.

Rotation periods for {\em bona fide} cluster members, 
in concert with photometric B-V colours, are used 
to create colour-period distributions, from which 
we outline a specific methodology for performing 
a gyrochronology analysis. In order to calculate 
the actual gyrochronology age of the cluster, we 
assign each M\,34 member its CgI status, as judged 
from its position in the colour-period diagram. 
I-sequence stars in M\,34 are classified specifically 
as those objects in the colour-period diagram lying 
above the I-sequence locus of the younger M\,35 
cluster, yielding a mean gyro age for all I-sequence 
stars in M\,34 of $193 \pm  9$ Myr.

There are two existing sets of age determination 
for the M\,34 cluster, using the traditional isochrone 
fitting method and the white dwarf cooling time-scale. 
Together with gyrochronology, all three methods suffer 
from various forms of dependency on stellar models 
under-pinning their theoretical frameworks. Furthermore, 
both the isochrone fitting and white dwarf methods suffer 
from a dependence on (or an assumption of) cluster 
distance, while gyrochronology is free from distance 
as an input parameter.

The isochronal age of M\,34 is not that well defined 
(see \S~\ref{extantM34} \& Table~\ref{table-age}), 
with a broad range in determined values. Unfortunately, 
the lack of observed giants in the cluster, and its broad 
{\em turn-off} locus in the V/B-V colour magnitude diagram 
precludes a precise assessment of its isochronal age. Based 
on existing studies, the mean isochronal age of the cluster 
is $250 \pm 67$ Myr. 

The white dwarf cooling time-scale age for M\,34 is also 
problematic because its result is based upon only five (5) 
objects whose membership of the cluster is not yet 
confirmed. Furthermore, the terminus of the white dwarf 
sequence has not yet been observed. In simple terms, the 
faintest, coolest, and therefore oldest white dwarfs have 
yet to be characterized, and the cluster's cooling 
time-scale age of $64.0 \pm 12.9$ Myr (Rubin et al. 2008) 
must be considered as a lower limit.

We conclude by noting that a study of M\,34 by Meibom et al., 
including both a multi-month differential photometry campaign 
and also a multi-year radial velocity membership and binarity 
campaign, similar to their prior work in M\,35, is in 
preparation. Their survey will yield several hundred new 
photometric periods of M\,34 solar-type stars, the I- and 
C-sequences of which should be far more clearly defined than 
they are from our Lowell campaign results. Their dataset should 
allow them to better define the gyrochronology age of M\,34, and 
we await their results eagerly.

\begin{acknowledgements}
This research has been supported by NSF grant 
AST-0349075 (Vanderbilt University), which is gratefully 
acknowledged. This research has made extensive use of 
the WEBDA database, operated by the Institute for 
Astronomy at the University of Vienna, as well as the 
SIMBAD database, operated at CDS, Strasbourg, 
France. Insightful discussions with Jonathon Irwin concerning 
his original M\,34 rotation period dataset are appreciated.
\end{acknowledgements}



%
%

\begin{table}
\caption{Extant isochronal ages for M\,34} 
\label{table-age}      
\centering                          
\begin{tabular}{rrr}        
\hline\hline                 
 Method    & Stellar Age & Reference \\
           &     [Myr]   &            \\
\hline                        
UBV       & 100     & Cester et al. (1977) \\
$uvby$    & 500     & Canterna et al. (1979) \\
UBV       & 250     & Ianna \& Schlemmer (1993) \\
UBV       & 180     & Meynet et al. (1993) \\
BVIc      & 200-250 & Jones \& Prosser (1996) \\
\hline                                   
\end{tabular}
\end{table}

\clearpage

\addtocounter{table}{1} 

\longtabL{2}{
\begin{landscape}
\begin{longtable}{rcclcrrrrrll}
\caption{\label{table-identifier} Astrometry, photometry and 
membership assignments are presented for those M\,34 
stars for which photometric periods are derived in 
the Lowell campaign.}\\
\hline\hline
Internal$^{a}$ & RA$^{b}$ & DEC$^{b}$ & JP 96$^{c}$ & 
V$^{d}$ &  Verr$^{d}$ &  B-V$^{d}$  & B-Verr$^{d}$ 
& V-I$^{d}$  & V-Ierr$^{d}$  & Phot$^{e}$ & RV$^{e}$ \\ 
Identifier    & (J2000) & (J2000) & & & & & & & & Mem & Mem \\
\hline
\endfirsthead
\caption{continued.}\\
\hline\hline
Internal & RA & DEC & JP 96 & V & Verr & B-V & B-Verr & V-I  
& V-Ierr  & Phot & RV \\ 
Identifier    & (J2000) & (J2000) & & & & & & & & Mem & Mem \\
\hline
\endhead
\hline
\endfoot
F3\_0172                  & 02 41 05.141 & +42 56 43.202 & JP 49  & 14.584 & 0.017 & 0.881 & 0.023 & 0.956 & 0.024 & Yes &  Yes-sing \\
F3\_0176                  & 02 41 36.611 & +42 50 20.288 & JP 167 & 14.659 & 0.009 & 0.888 & 0.013 & 1.051 & 0.013 & Yes &  Yes-sing \\
F3\_0215                  & 02 41 00.981 & +42 52 46.695 & JP 41  & 14.918 & 0.010 & 0.944 & 0.015 & 1.039 & 0.014 & Yes &  Yes-sing \\
F3\_0258                  & 02 42 02.552 & +42 51 51.721 & JP 289 & 15.236 & 0.009 & 1.017 & 0.013 & 1.118 & 0.013 & Yes &  Yes-sing \\
F3\_0306                  & 02 41 57.472 & +43 00 26.530 & JP 265 & 15.503 & 0.014 & 1.061 & 0.019 & 1.262 & 0.027 & Yes &  Yes-bin  \\
F3\_0320                  & 02 41 37.557 & +42 57 21.593 & JP 172 & 15.664 & 0.017 & 1.116 & 0.025 & 1.237 & 0.024 & Yes &           \\
F3\_0383                  & 02 40 49.882 & +42 55 31.520 &        & 16.179 & 0.017 & 1.230 & 0.025 & 1.424 & 0.024 & Yes &  Yes-sing \\
F3\_0430                  & 02 41 50.490 & +42 58 10.016 &        & 16.424 & 0.013 & 1.309 & 0.019 & 1.505 & 0.018 & Yes &  Yes-sing \\
F3\_0469                  & 02 40 49.090 & +42 48 20.648 &        & 16.604 & 0.018 & 1.353 & 0.030 & 1.550 & 0.025 & Yes &           \\
F3\_0485                  & 02 41 06.638 & +42 48 22.011 &        & 16.732 & 0.010 & 1.355 & 0.016 & 1.619 & 0.014 & Yes &           \\
F3\_0487                  & 02 40 44.889 & +42 53 47.221 &        & 16.818 & 0.020 & 1.348 & 0.030 & 1.701 & 0.027 & Yes &           \\
F3\_0600                  & 02 41 18.447 & +42 58 21.342 &        & 17.394 & 0.020 & 1.522 & 0.035 & 1.863 & 0.027 & Yes &           \\
F3\_0664                  & 02 41 20.454 & +42 58 52.278 &        & 17.862 & 0.022 & 1.511 & 0.057 & 1.999 & 0.029 & Yes &           \\
F4\_0136                  & 02 41 47.384 & +42 43 38.525 & JP 213 & 13.151 & 0.010 & 0.614 & 0.014 & 0.718 & 0.015 & Yes &  Yes-sing \\
F4\_0169                  & 02 41 04.952 & +42 46 51.343 & JP 50  & 13.510 & 0.010 & 0.668 & 0.014 & 0.786 & 0.014 & Yes &  Yes-sing \\
F4\_0194                  & 02 41 28.142 & +42 38 37.095 & JP 133 & 13.662 & 0.010 & 0.722 & 0.014 & 0.805 & 0.015 & Yes &  Yes-sing \\
F4\_0234                  & 02 41 33.429 & +42 42 11.625 & JP 148 & 13.959 & 0.010 & 0.768 & 0.014 & 0.844 & 0.014 & Yes &  Yes-sing \\
F4\_0303                  & 02 41 49.932 & +42 37 13.906 & JP 227 & 14.366 & 0.009 & 0.837 & 0.013 & 0.912 & 0.013 & Yes &  Yes-sing \\
F4\_0317                  & 02 41 49.354 & +42 36 37.613 & JP 224 & 14.410 & 0.010 & 0.875 & 0.014 & 0.959 & 0.014 & Yes &  Yes-sing \\
F4\_0327                  & 02 41 44.167 & +42 46 07.333 & JP 199 & 14.483 & 0.009 & 0.883 & 0.013 & 0.941 & 0.013 & Yes &  Yes-sing \\
F4\_0335                  & 02 41 35.247 & +42 41 02.323 & JP 158 & 14.546 & 0.010 & 0.860 & 0.014 & 0.994 & 0.014 & Yes &  Yes-bin  \\
F4\_0450                  & 02 41 23.098 & +42 40 15.954 & JP 113 & 14.960 & 0.010 & 0.965 & 0.014 & 1.058 & 0.014 & Yes &  Yes-sing \\
F4\_0667                  & 02 41 06.187 & +42 46 20.783 & JP 52  & 15.623 & 0.009 & 1.100 & 0.013 & 1.233 & 0.013 & Yes &  Yes-bin  \\
F4\_0695                  & 02 41 42.922 & +42 33 13.910 & JP 197 & 15.643 & 0.021 & 1.156 & 0.030 & 1.289 & 0.029 & Yes &  Yes-sing \\
F4\_0730                  & 02 40 49.647 & +42 46 54.799 & JP 18  & 15.741 & 0.017 & 1.121 & 0.025 & 1.257 & 0.024 & Yes &           \\
F4\_0736                  & 02 40 48.542 & +42 39 25.716 &        & 15.844 & 0.021 & 1.078 & 0.028 & 1.328 & 0.029 & Yes &  Yes-bin  \\
F4\_0803                  & 02 41 50.301 & +42 44 37.371 & JP 229 & 16.025 & 0.010 & 1.234 & 0.014 & 1.373 & 0.013 & Yes &  Yes-sing \\
F4\_0804                  & 02 41 48.974 & +42 39 59.680 &        & 16.034 & 0.010 & 1.176 & 0.015 & 1.378 & 0.014 & Yes &           \\
F4\_0852                  & 02 41 53.246 & +42 35 26.394 &        & 16.294 & 0.009 & 1.225 & 0.013 & 1.634 & 0.013 & Yes &  Yes-bin  \\
F4\_0942                  & 02 41 46.371 & +42 32 32.128 &        & 16.341 & 0.020 & 1.268 & 0.029 & 1.498 & 0.028 & Yes &           \\
F4\_1007                  & 02 40 50.752 & +42 41 19.844 &        & 16.839 & 0.022 & 1.510 & 0.033 & 1.901 & 0.034 & Yes &           \\
F4\_1020                  & 02 40 36.693 & +42 41 32.500 &        & 16.608 & 0.021 & 1.330 & 0.031 & 1.584 & 0.030 & Yes &           \\
F4\_1055                  & 02 40 42.791 & +42 38 59.153 &        & 16.639 & 0.021 & 1.342 & 0.031 & 1.600 & 0.029 & Yes &           \\
F4\_1147                  & 02 41 43.854 & +42 45 07.859 &        & 17.009 & 0.010 & 1.406 & 0.017 & 1.749 & 0.013 & Yes &           \\
F4\_1315                  & 02 40 53.997 & +42 38 48.513 &        & 17.347 & 0.022 & 1.441 & 0.036 & 1.815 & 0.030 & Yes &           \\
F4\_1357                  & 02 41 44.198 & +42 35 36.074 &        & 17.476 & 0.011 & 1.450 & 0.021 & 1.911 & 0.015 & Yes &           \\
F4\_1369                  & 02 41 51.600 & +42 29 50.864 &        & 17.496 & 0.015 & 1.513 & 0.027 & 1.934 & 0.021 & Yes &           \\
F4\_1392                  & 02 41 38.851 & +42 43 01.148 &        & 17.654 & 0.011 & 1.504 & 0.025 & 1.963 & 0.015 & Yes &           \\
F4\_1566                  & 02 41 31.131 & +42 41 14.283 &        & 18.132 & 0.012 & 1.549 & 0.032 & 2.137 & 0.016 & Yes &           \\
F4\_1617                  & 02 41 26.292 & +42 30 15.124 &        & 18.165 & 0.027 & 1.570 & 0.064 & 2.178 & 0.035 & Yes &           \\
F4\_1793                  & 02 41 20.042 & +42 39 23.652 &        & 18.606 & 0.015 & 1.393 & 0.063 & 2.285 & 0.019 & Yes &           \\
F4\_1925$^{f}$            & 02 41 21.947 & +42 32 32.637 &        & 19.040 & 0.037 & 1.577 & 0.045 & 2.350 & 0.045 & Yes &          \\
F4\_2226$^{f}$            & 02 40 31.108 & +42 40 25.105 &        & 19.659 & 0.067 & 1.625 & 0.071 & 2.626 & 0.071 & Yes &           \\ \hline
F3\_0071                  & 02 40 44.422 & +42 54 04.516 &        & 12.855 & 0.019 & 1.047 & 0.025 & 1.153 & 0.026 &  No &           \\
F3\_0121                  & 02 41 47.414 & +42 50 09.223 & JP 212 & 13.741 & 0.009 & 1.248 & 0.013 & 1.354 & 0.013 &  No &           \\
F3\_0163                  & 02 41 55.003 & +42 52 53.485 & JP 258 & 14.440 & 0.008 & 0.900 & 0.011 & 0.986 & 0.012 & Yes &  No       \\
F3\_0165                  & 02 41 51.794 & +42 46 20.473 & JP 244 & 14.419 & 0.008 & 0.865 & 0.011 & 0.956 & 0.011 & Yes &  No       \\
F3\_0388                  & 02 41 38.227 & +42 55 43.112 & JP 174 & 15.952 & 0.010 & 0.975 & 0.015 & 1.076 & 0.022 &  No &           \\
F3\_0413                  & 02 41 48.478 & +42 49 33.840 & JP 218 & 16.242 & 0.010 & 1.143 & 0.015 & 1.412 & 0.014 & Yes &  No       \\
F3\_0437                  & 02 40 48.970 & +42 56 12.434 &        & 16.022 & 0.018 & 0.647 & 0.025 & 0.748 & 0.025 &  No &           \\
F3\_0692                  & 02 40 57.994 & +42 49 44.614 &        & 17.136 & 0.010 & 0.857 & 0.016 & 1.061 & 0.015 &  No &           \\
F3\_2310                  & 02 40 44.085 & +42 53 58.059 &        & 99.999 & 9.999 & 9.999 & 9.999 & 9.999 & 9.999 &     &           \\
F4\_0167                  & 02 41 48.825 & +42 31 31.702 & JP 223 & 13.502 & 0.021 & 0.938 & 0.026 & 1.030 & 0.031 &  No &           \\
F4\_0404                  & 02 41 27.663 & +42 35 42.306 & JP 131 & 14.730 & 0.010 & 0.852 & 0.014 & 0.932 & 0.014 & Yes &  No       \\
F4\_0515                  & 02 41 27.188 & +42 43 41.881 & JP 129 & 15.111 & 0.010 & 0.903 & 0.014 & 0.977 & 0.014 &  No &           \\
\end{longtable}
\noindent $a -$ Identifier format is XX\_YYYY, where 'XX' are 
representations F3 and F4 for targets detected on M\,34 
western Field 3 or western Field 4 images, \\
\noindent and 'YYYY' are internal catalog numbers.\\
\noindent $b -$ Astrometric data are derived from coordinate 
matches in the SuperCosmos survey catalogs (Hambly et al. 
2001a, b, c). \\
\noindent $c -$ JP 96 identifiers are for photometric variables which are 
correlated with Jones \& Prosser (1996) sources \\
\noindent $d -$ BVI photometric data are taken from the 
Indiana Group survey (\cf \S~\ref{indiana}).\\
\noindent $e -$ Membership assignments are based upon whether 
a target star lies on (or close to) the V/B-V and/or V/V-I 
main sequence loci [see Fig.,~\ref{CMDs}] or whether a 
target star has a radial velocity consistent with the 
cluster's systemic velocity (Meibom et al. in prep; see 
also \S~\ref{extantOBS} and Meibom et al. 2006). \\
\noindent $f -$ B-V colours are determined using a field 
star V-I to B-V relationship (Caldwell et al. 1993), 
assuming E$_{V-I}$ $= 1.25\times$E$_{B-V}$, and E$_{B-V}$=0.07 
(Canterna et al. 1979). \\
\end{landscape}
}

\clearpage

\longtab{3}{
\begin{longtable}{lrrrrl}
\caption{\label{table-Prot} Photometric periods and 
ancillary data are presented, resulting from the light-curve 
analysis of variable stars detected in the Lowell M\,34 fields.}\\
\hline\hline
Internal$^{a}$  ~~    & Data$^{b}$ & Sigma Clipped$^{c}$ & Period 
                    & B-V$_{o}^{d}$ & Mem$^{e}$ \\
Identifier  & Epochs & StdDev [mags] & [days] & &  \\
\hline
\endfirsthead
\caption{continued.}\\
\hline\hline
Internal    & Data   & Sigma Clipped & Period & B-V$_{o}$ & Mem \\
Identifier  & Epochs & StdDev [mags] & [days] &           &     \\
\hline
\endhead
\hline
\endfoot
  F3\_0172-lIlVsV   & 53/54/59     & 0.00883/0.01010/0.01190          & $ 8.0    \pm  1.0  $ & 0.811 & Y;Ys \\
  F3\_0176-lIlVsIsV & 53/54/58/59  & 0.04275/0.06460/0.04313/0.06427  & $ 2.780  \pm  0.005$ & 0.818 & Y;Ys \\
  F3\_0215-lIlVsIsV & 52/54/58/59  & 0.01121/0.01722/0.01439/0.02020  & $ 7.47   \pm  0.11 $ & 0.874 & Y;Ys \\
  F3\_0258-lIlVsV   & 53/54/58     & 0.01182/0.01250/0.01457          & $11.0    \pm  1.0  $ & 0.947 & Y;Ys \\
  F3\_0306-lIsIsV   & 53/58/59     & 0.01886/0.01809/0.02081          & $1.1435  \pm 0.0007$ & 0.991 & Y;Yb \\
  F3\_0320-lIlVsV   & 52/54/58     & 0.00898/0.01023/0.01430          & $ 8.511  \pm 0.001 $ & 1.046 & Y;   \\
  F3\_0383-lIlV     & 52/54        & 0.00776/0.00937                  & $ 9.5    \pm  0.4  $ & 1.160 & Y;Ys \\
  F3\_0430-lV       & 54           & 0.01253                          & $10.5    \pm 1.0   $ & 1.239 & Y;Ys \\
  F3\_0469-lIlVsIsV & 52/54/57/57  & 0.01344/0.01745/0.01768/0.02225  & $12.9    \pm  0.3  $ & 1.283 & Y; \\
  F3\_0485-lIlV     & 52/54        & 0.01040/0.01393                  & $12.0    \pm  1.0  $ & 1.285 & Y; \\
  F3\_0487-lIlV     & 53/54        & 0.01076/0.01388                  & $ 8.4    \pm  0.6  $ & 1.278 & Y; \\
  F3\_0600-lIlV     & 53/54        & 0.01263/0.02503                  & $ 7.44   \pm  0.05 $ & 1.452 & Y; \\
  F3\_0664-lI       & 52           & 0.01077                          & $10.5    \pm 1.0   $ & 1.441 & Y; \\
  F4\_0136-sIsV     & 60/60        & 0.01457/0.01794                  & $ 2.87   \pm  0.03 $ & 0.544 & Y;Ys \\ 
  F4\_0169-lIlVsIsV & 55/42/60/59  & 0.01275/0.01297/0.01094/0.01429  & $ 3.81   \pm  0.08 $ & 0.598 & Y;Ys \\ 
  F4\_0194-sIsV     & 60/60        & 0.00817/0.00889                  & $ 2.20   \pm  0.03 $ & 0.652 & Y;Ys \\ 
  F4\_0234-lIlVsV   & 54/55/60     & 0.00692/0.00984/0.01014          & $ 5.56   \pm  0.07 $ & 0.698 & Y;Ys \\ 
  F4\_0303-lIlVsIsV & 55/55/60/60  & 0.00800/0.00941/0.00839/0.00978  & $ 7.4    \pm  0.3  $ & 0.767 & Y;Ys \\ 
  F4\_0317-lIlVsIsV & 55/55/59/60  & 0.01701/0.02159/0.01698/0.02167  & $ 6.8    \pm  0.1  $ & 0.805 & Y;Ys \\ 
  F4\_0327-lVsV     & 55/60        & 0.01471/0.01406                  & $ 7.6    \pm  1.1  $ & 0.813 & Y;Ys \\ 
  F4\_0335-lIlVsIsV & 55/55/60/60  & 0.01538/0.01953/0.01601/0.02032  & $ 0.89   \pm  0.02 $ & 0.790 & Y;Yb \\ 
  F4\_0450-lIlVsIsV & 54/55/60/60  & 0.00985/0.01181/0.01264/0.01368  & $ 8.00   \pm  0.21 $ & 0.895 & Y;Ys \\ 
  F4\_0667-lIlVsV   & 54/55/59     & 0.01122/0.01819/0.02028          & $ 7.97   \pm  0.05 $ & 1.030 & Y;Yb \\ 
  F4\_0695-lI       & 54           & 0.00748                          & $ 4.3    \pm 0.5   $ & 1.086 & Y;Ys \\
  F4\_0730-lIlVsIsV & 55/55/60/59  & 0.02944/0.04619/0.03109/0.04695  & $ 8.41   \pm  0.10 $ & 1.051 & Y; \\
  F4\_0736-lIlVsIsV & 55/55/60/60  & 0.01875/0.02715/0.02022/0.02911  & $ 0.786  \pm  0.001$ & 1.008 & Y;Yb \\ 
  F4\_0803-lVsI     & 55/60        & 0.01435/0.01416                  & $12.6    \pm  0.7  $ & 1.164 & Y;Ys \\ 
  F4\_0804-lIlVsIsV & 55/55/60/60  & 0.02568/0.04152/0.02448/0.04239  & $ 4.95   \pm  0.04 $ & 1.106 & Y; \\
  F4\_0852-lIlVsIsV & 55/55/59/60  & 0.01547/0.02368/0.01624/0.02564  & $0.879   \pm  0.002$ & 1.155 & Y;Yb \\ 
  F4\_0942-lIlVsV   & 54/55/60     & 0.00942/0.01300/0.02362          & $ 9.4    \pm  0.3  $ & 1.198 & Y; \\
  F4\_1007-lIlV     & 55/55        & 0.00774/0.01177                  & $15      \pm 2     $ & 1.440 & Y; \\ 
  F4\_1020-lIlVsIsV & 55/55/60/59  & 0.01344/0.01846/0.01568/0.02528  & $ 8.6    \pm  0.4  $ & 1.260 & Y; \\
  F4\_1055-lIlVsIsV & 55/55/60/60  & 0.01427/0.02542/0.02233/0.02664  & $ 6.80   \pm  0.15 $ & 1.272 & Y; \\
  F4\_1147-lV       & 55           & 0.01883                          & $ 0.589  \pm 0.001 $ & 1.336 & Y; \\
  F4\_1315-lIlV     & 54/54        & 0.01429/0.01913                  & $12.2    \pm  0.5  $ & 1.371 & Y; \\
  F4\_1357-lIlV     & 55/55        & 0.01197/0.01973                  & $ 9.4    \pm  0.2  $ & 1.380 & Y; \\
  F4\_1369-lIlV     & 53/53        & 0.01886/0.02578                  & $11.85   \pm  0.13 $ & 1.443 & Y; \\
  F4\_1392-lIlV     & 55/55        & 0.01496/0.02385                  & $ 6.5    \pm  0.5  $ & 1.434 & Y; \\
  F4\_1566-lIlV     & 55/54        & 0.01920/0.03248                  & $11.2    \pm  0.1  $ & 1.479 & Y; \\
  F4\_1617-lV       & 52           & 0.03379                          & $ 7.3    \pm 0.5   $ & 1.500 & Y;  \\
  F4\_1793-lIlV     & 55/53        & 0.02221/0.04966                  & $ 7.5    \pm  0.6  $ & 1.323 & Y; \\
  F4\_1925-lV$^{f}$ & 44           & 0.09210                          & $12.7    \pm 2.0   $ & 1.507 & Y;  \\
  F4\_2226-lI$^{f}$ & 55           & 0.02681                          & $ 6.0    \pm 0.5   $ & 1.555 & Y;  \\ \hline
  F3\_0071-sIsV     & 58/59        & 0.02052/0.02502                  & $ 3.014  \pm 0.008 $ & 0.977 & N;  \\
  F3\_0121-lVsV     & 54/59        & 0.00851/0.00881                  & $ 8.4    \pm 0.2   $ & 1.178 & N; \\
  F3\_0163-lIsIsV   & 53/58/59     & 0.00814/0.00943/0.00935          & $14      \pm  2    $ & 0.830 & Y;N \\
  F3\_0165-lI       & 52           & 0.01089                          & $12      \pm 1     $ & 0.795 & Y;N \\
  F3\_0388-lIlVsIsV & 53/54/58/59  & 0.01534/0.02141/0.01715/0.02448  & $11.59   \pm 0.01  $ & 0.905 & N; \\
  F3\_0413-lIlVsIsV & 53/54/58/59  & 0.01640/0.02417/0.01909/0.02982  & $ 5.45   \pm  0.08 $ & 1.073 & Y;N \\
  F3\_0437-lIlVsV   & 53/54/59     & 0.00859/0.01134/0.01563          & $ 9.7    \pm 0.4   $ & 0.577 & N; \\
  F3\_0692-lIlV     & 53/54        & 0.01886/0.02686                  & $ 0.3368 \pm 0.0001$ & 0.787 & N; \\
  F3\_2310-lIlV     & 52/54        & 0.02667/0.02846                  & $ 2.991  \pm 0.008 $ & - - - & \\
  F4\_0167-lIlVsI   & 54/55/59     & 0.01321/0.01091/0.00961          & $ 9.2    \pm 0.9   $ & 0.868 & N; \\
  F4\_0404-lIlVsV   & 55/55/60     & 0.01099/0.01419/0.01461          & $ 6.6    \pm  0.3  $ & 0.782 & Y;N \\
  F4\_0515-lIlVsV   & 55/55/60     & 0.00886/0.00957/0.01239          & $ 8.2    \pm 0.1   $ & 0.833 & N; 
\end{longtable}
\noindent $a -$ Internal Identifiers are the same as those detailed in 
Table~\ref{table-identifier}. Designations indicating exposure time and filter 
choice for a given light-curve are annotated. For instance, {\em lI} represents 
{\em long-I}, for light-curves derived from long exposures in the I-filter. 
Thus, {\em lIlVsIsV} represents four independent light-curves created from 
image sequences of 400seconds, I-filter; 600seconds, V-filter; 40seconds, 
I-filter \& 60seconds, V-filter, respectively (namely, long-I, long-V, 
short-I \& short-V).\\
\noindent $b -$ Number of photometric data points in each light-curve, 
derived from each target's image series, in a given filter and exposure 
time sequence. \\
\noindent $c -$ For each target's light-curve, in a filter+exposure 
combination, standard deviations of the differential magnitudes are 
provided. The time series files for each target were sigma-clipped 
to eliminate outliers. A $\Delta$-magnitude for each light-curve can 
be determined by multiplying these standard deviations by 2$\sqrt2$ 
(strictly speaking this is only true for perfectly sinusoidal 
variations). \\
\noindent $d -$ E$_{B-V}$=0.07 is adopted; Canterna et al. (1979) \\
\noindent $e -$ Abridged cluster membership criteria are reproduced 
from Table~\ref{table-identifier}, with format Photometric; 
Kinematic (s-single, b-binary). \\
\noindent $f -$ B-V colours are determined using a field 
star V-I to B-V relationship (Caldwell et al. 1993), assuming 
E$_{V-I}$ $= 1.25\times$E$_{B-V}$, and E$_{B-V}$=0.07.}

\clearpage

\begin{table}
\caption{A comparison of photometric periods for 
three M\,34 stars common to the Lowell campaign, 
the Irwin et al. study and the Meibom et al. (in prep) study 
(\cf Fig.,~\ref{IRWINcompFIG} and ~\S~\ref{IRWINcomp}).}
\label{soren-syd-irwin-3stars}
\centering                          
\begin{tabular}{cccc}        
\hline\hline                 

ID           &  Lowell                & Irwin et al. & Meibom et al.  \\
             &  [days]                &  [days]      & [days] \\
\hline
F3\_0413     & $5.45  \pm 0.08$       & 2.458        & 5.47  \\
F4\_0234     & $5.56  \pm 0.07$       & 0.822        & 5.47  \\
F4\_1147     & $0.589 \pm 0.001$      & 1.429        & 0.592 \\
\hline
\end{tabular}
\end{table}

\clearpage

%
%
\begin{table*}
\caption{Gyrochronological ages for I-sequence stars in M\,34}
\label{Istars-table-Prot}
\begin{tabular}{clccllll}
\hline\hline
Identifier  & ~~~JP$^{a}$ & $\#$ of            & (B-V)o & ~Measured         & ~~Gyro Age$^{c}$   & Fitted$^{d}$  & Period$^{e}$       \\
            & ~~~ no.     & periods            &        & ~~~Period         &                    & Period        & Variance     \\
            &             & derived$^{b}$      &        & ~~~~[days]       & ~~~~~[Myr]         & [days]        & [days$^{2}$] \\
\hline
  F4\_0136  & JP 213      &   2                & 0.544  & $ 2.87  \pm 0.03$  & $178.5  \pm   3.5$ &  ~2.99        & 0.0144  \\
  F4\_0169  & JP 50       &   4                & 0.598  & $ 3.81  \pm 0.08$  & $170.0  \pm   6.7$ &  ~4.08        & 0.0729  \\
  F4\_0234  & JP 148      &   3                & 0.698  & $ 5.56  \pm 0.07$  & $188.4  \pm   4.4$ &  ~5.63        & 0.0049  \\
  F4\_0303  & JP 227      &   4                & 0.767  & $ 7.4   \pm 0.3$   & $244.1  \pm  18.6$ &  ~6.5         & 0.81    \\
  F4\_0317  & JP 224      &   4                & 0.805  & $ 6.8   \pm 0.1$   & $183.8  \pm   5.0$ &  ~7.0         & 0.04    \\
  F4\_0327  & JP 199      &   2                & 0.813  & $ 7.6   \pm 1.1$   & $220.9  \pm  59.8$ &  ~7.1         & 0.25    \\
  F3\_0172  & JP 49       &   3                & 0.811  & $ 8.0   \pm 1.0$   & $244.6  \pm  57.2$ &  ~7.0         & 1.00    \\
  F3\_0215  & JP 41       &   4                & 0.874  & $ 7.47  \pm 0.11$  & $180.4  \pm   5.0$ &  ~7.74        & 0.0729  \\
  F4\_0450  & JP 113      &   4                & 0.895  & $ 8.00  \pm 0.21$  & $194.5  \pm   9.6$ &  ~7.96        & 0.0016  \\
  F3\_0258  & JP 289      &   3                & 0.947  & $11.0   \pm 1.0$   & $313.1  \pm  53.3$ &  ~8.5         & 6.25    \\
  F4\_0667  & JP 52       &   3                & 1.030  & $ 7.97  \pm 0.05$  & $145.0  \pm   1.7$ &  ~9.28        & 1.7161  \\
  F3\_0320  & JP 172      &   3                & 1.046  & $ 8.511 \pm 0.001$ & $159.3  \pm   0.1$ &  ~9.429       & 0.84272 \\
  F4\_0730  & JP 18       &   4                & 1.051  & $ 8.41  \pm 0.10$  & $154.4  \pm   3.4$ &  ~9.47        & 1.1236  \\
  F3\_0383  &             &   2                & 1.160  & $ 9.5   \pm 0.4$   & $162.2  \pm  12.8$ &  10.4         & 0.81    \\
  F4\_0803  & JP 229      &   2                & 1.164  & $12.6   \pm 0.7$   & $273.5  \pm  28.5$ &  10.5         & 4.41    \\
  F4\_0942  &             &   3                & 1.198  & $ 9.4   \pm 0.3$   & $150.4  \pm   9.0$ &  10.7         & 1.69    \\
  F3\_0430  &             &   1                & 1.239  & $10.5   \pm 1.0$   & $174.8  \pm  31.2$ &  11.1         & 0.36    \\
  F3\_0485  &             &   2                & 1.285  & $12.0   \pm 1.0$   & $211.3  \pm  33.0$ &  11.4         & 0.36    \\
  F3\_0469  &             &   4                & 1.283  & $12.9   \pm 0.3$   & $242.5  \pm  10.6$ &  11.4         & 2.25    \\
  F4\_1315  &             &   2                & 1.371  & $12.2   \pm 0.5$   & $196.4  \pm  15.1$ &  12.1         & 0.01    \\
  F4\_1357  &             &   2                & 1.380  & $ 9.4   \pm 0.2$   & $119.3  \pm   4.8$ &  12.2         & 7.84    \\ 
  F3\_0664  &             &   1                & 1.441  & $10.5   \pm 1.0$   & $137.2  \pm  24.5$ &  12.6         & 4.41    \\
  F4\_1369  &             &   2                & 1.443  & $11.85  \pm 0.13$  & $171.7  \pm   3.5$ &  12.61        & 0.5776  \\
  F4\_1566  &             &   2                & 1.479  & $11.2   \pm 0.1$   & $148.8  \pm   2.5$ &  12.9         & 2.89    \\
  F4\_1007  &             &   2                & 1.440  & $15.0   \pm 2.0$   & $267.8  \pm  66.8$ &  12.6         & 5.76    \\
  F4\_1925  &             &   1                & 1.507  & $12.7   \pm 2.0$   & $183.0  \pm  53.9$ &  13.1         & 0.16    \\ \hline
\end{tabular} \\

\noindent $a -$ {\em JP} identifier from Jones \& Prosser (1996).\\
\noindent $b -$ Number of filter/exposure combinations yielding 
photometric period derivations for a given star.\\
\noindent $c -$ Gyro ages determined using Meibom et al. (2009) 
I-sequence colour function coefficients, {\em viz} $a=0.770 \pm 0.014$, 
$b=0.553 \pm 0.052$ \& $c=0.472\pm0.027$, and an age function 
exponent $n=0.5344\pm0.0015$ (see \S~\ref{M34gyro}).\\
\noindent $d -$ Assuming a mean gyrochronology age for M\,34 of 193Myr 
(the mean of the ages calculated in column 6 - see \S~\ref{M34gyro}) 
and the Meibom et al. (2009) coefficients, with $n=0.5344\pm0.0015$, we 
use intrinsic (B-V) colours and Equ.,1 to derive a fitted period 
for each M\,34 I-sequence star.\\
\noindent $e -$ Each star's contribution to the period variance 
[(Measured Period - Fitted Period)$^{2}$] is noted, where the total 
variance is 43.73 days${^2}$, with a mean of 1.68 days${^2}$.\\
\end{table*}

\clearpage

%
%

\begin{figure*}
\includegraphics[angle=-90,scale=0.65]{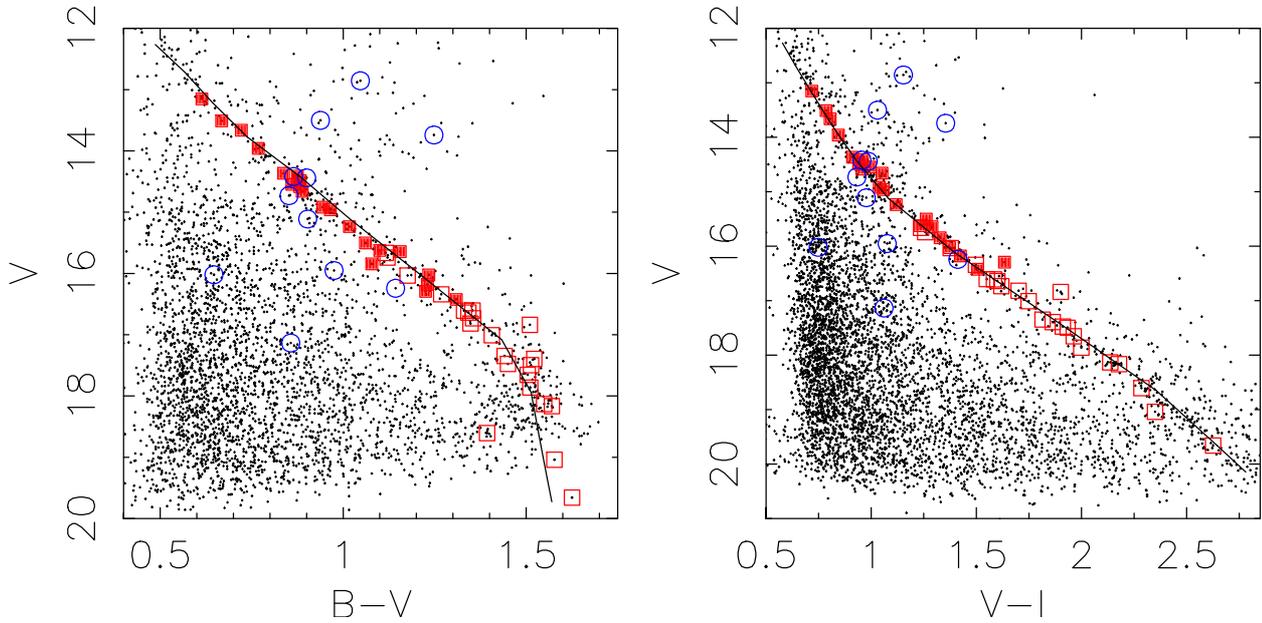}
\caption{Colour magnitude diagrams, in B-V and V-I colours, 
are presented for the entire photometry catalogue of the 
M\,34 BVIc survey conducted by the Indiana Group. 
Solid lines represent D'Antona \& Mazzitelli (1997) 
theoretical isochrones, for a cluster age of 250 Myr, a 
distance of 470pc, and assuming E$_{V-I}$ $= 1.25\times$ 
E$_{B-V}$, and E$_{B-V}$=0.07 (Canterna et al. 1979). 
Red squares, highlighting photometric and radial velocity 
members of the cluster (solid symbols), and photometric cluster 
membership only (open symbols) are plotted for M\,34 stars for 
which we have derived photometric periods during the Lowell 
campaign. Blue circles depict those other periodic photometric 
variables in the field of M\,34 which are photometric and/or 
radial velocity non-members of the cluster.}
\label{CMDs}
\end{figure*}

\begin{figure*}
\includegraphics[angle=-90,scale=0.65]{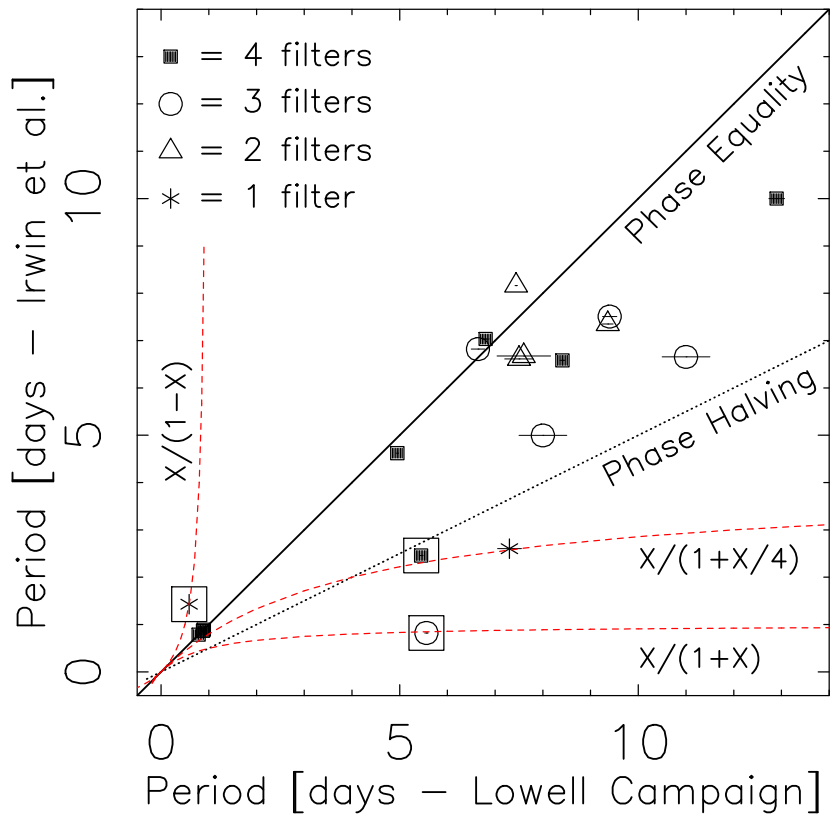}
\caption{Photometric periods for M\,34 stars common to our Lowell 
campaign and the Irwin et al. study are shown. Periods determined 
from the Lowell campaign, using photometric light-curves in up 
to four (4) different filter/exposure time combinations, are 
depicted by the filled-square, circle, triangle and asterisk 
symbols (see Tables~\ref{table-identifier} \& ~\ref{table-Prot}). 
We include error bars on our Lowell periods, which for many 
targets are smaller than the symbol sizes. No error bars are 
available for the Irwin et al. periods. The solid line represents 
equality between the two studies, and is not a fit to the data. The 
dotted line represents phase-halving while the dashed lines represent 
several period-aliasing loci, respectively. About one-third of the sample 
have equal period determinations (within the errors). Of the remainder, 
all but two of the Irwin et al. periods are shorter than ours. See text 
for possible reasons for the discrepancy (see \S~\ref{IRWINcomp}). 
The three flagged stars [with boxed symbols] are discrepant ones where 
our periods have been confirmed by a forthcoming Meibom et al. 
(in prep) study.}
\label{IRWINcompFIG}
\end{figure*}

\begin{figure*}
\includegraphics[angle=-90,scale=0.65]{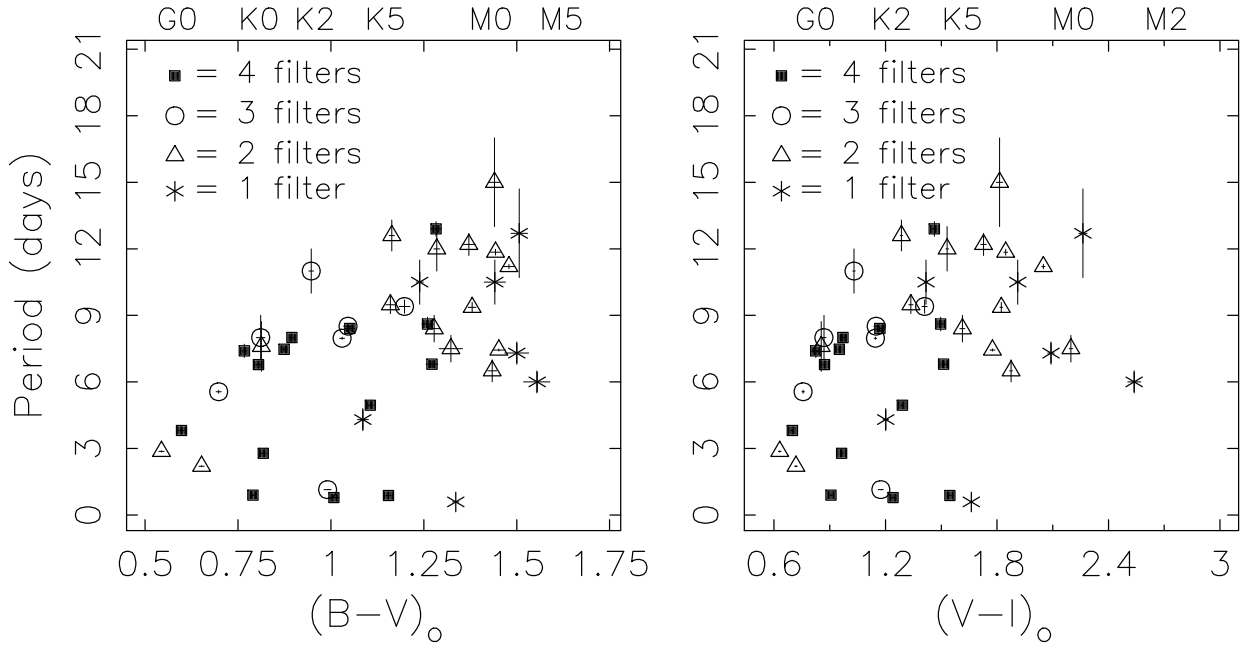}
\caption{Photometric period distributions, as a function of 
intrinsic colours (B-V)$_{o}$ and (V-I)$_{o}$, are plotted 
for solar-type stars in NGC 1039 (M\,34). We adopt 
E$_{B-V}$=0.07 (Canterna et al. 1979), and 
E$_{V-I}$ $= 1.25\times$ E$_{B-V}$. Period symbols are 
the same as those depicted in Fig.,~\ref{IRWINcompFIG}. 
Error bars are included, and are for some targets smaller 
than the symbol sizes. Approximate spectral 
types for a given colour are annotated on the upper 
abscissae (from Zombeck 1990).}
\label{M34-Prot-BVoVIo}
\end{figure*}

\clearpage

\clearpage

\Online

\begin{appendix} 

\section{Cluster Membership of M\,34 Using Extant Kinematic Measurements}\label{J97mem}

There exist two extant kinematic studies of the M\,34 cluster which can assist 
us further in establishing cluster membership for our gyrochronology sample. The 
first, by Jones \& Prosser (1996), relies upon 2-d space motions in the form of 
proper motions, although their high-fidelity membership probabilities are magnitude 
limited to about V$\simeq 14.5$. While JP96 proper motion results do extend to 
fainter magnitudes, their reliability reduces. That is to say, a fainter M\,34 
star with a $00\%$ proper motion probability may indeed still be a {\em bona fide} 
member of the cluster, insomuch as Poisson-limited proper motion errors could 
lead to erroneously lower proper motion probabilities.

The second, by Jones et al. (1997), employing 1-d heliocentric radial velocities, 
can be used to establish cluster membership, or in the very least, establish 
cluster non-membership (or binarity). Unfortunately neither membership probabilities 
nor velocity errors were reported by Jones et al., however we can exploit their 
measurements in our membership assessment (see below). Interestingly, in the same 
manuscript, Jones et al. report \lii 6708\r{A} equivalent widths [EWs] for their 
target stars. Even though the detection of lithium in solar type stars is not 
a requisite for cluster membership, its absence is almost ubiquitous among the 
generally older, Galactic field stars. The very presence of a substantial lithium 
line in solar-type stellar spectra is in itself indicative of youth 
($<<1000$ Myr; {\em e.g.,} Soderblom et al. 1993b, James \& Jeffries 1997, 
James et al. 2006).

\begin{figure*}
\includegraphics[angle=-90,width=16.4cm,clip]{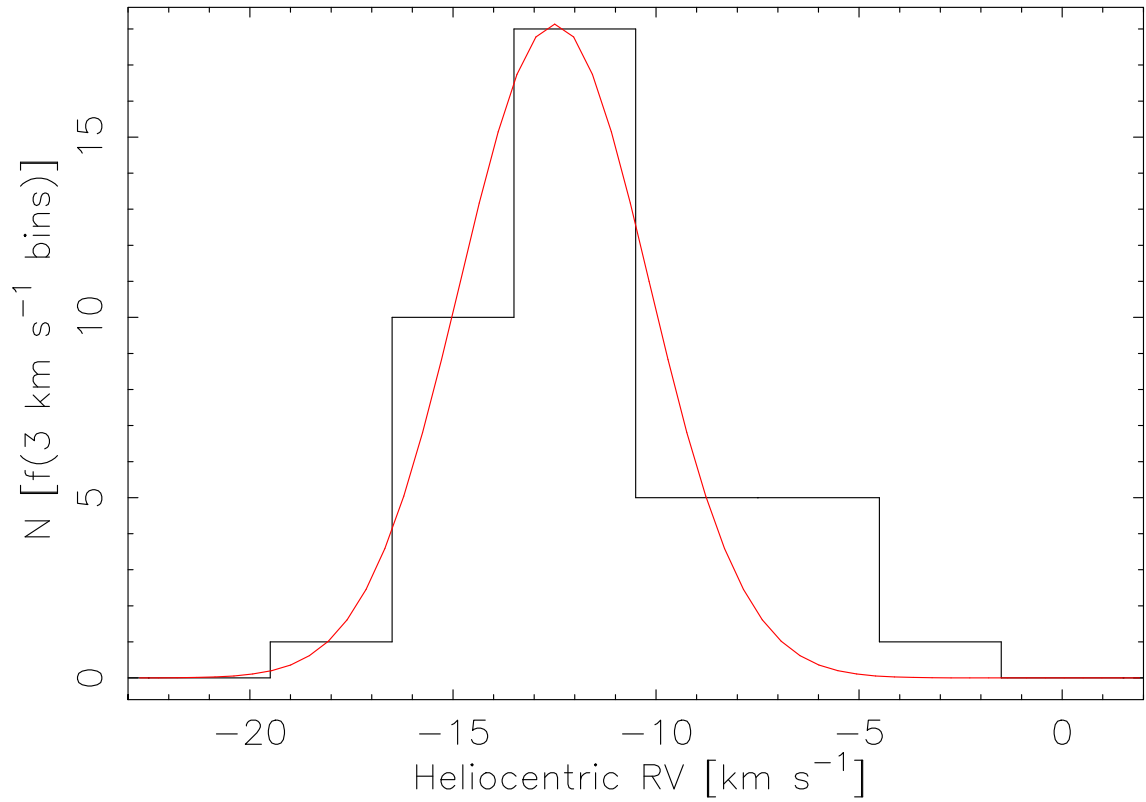}
\caption{Heliocentric radial velocities, in histogram form using 3~km~s$^{-1}$ 
bins, are plotted for M\,34 stars using velocity data detailed in Jones et al. 
(1997). The red solid line depicts an unweighted Gaussian fit to the data, with 
a Gaussian centre of -12.49 km s$^{-1}$ and a sigma of 2.32 km s$^{-1}$. }
\label{J97-RV-gauss}
\end{figure*}

For each of the M\,34 stars in our sample, we have collated these additional 
membership criteria published by JP96 and Jones et al. (1997) in 
Table~\ref{GYRO-vs-Jones97}. Its Lowell identifier is listed in column~1 in 
concert with its JP96 identifier in column~2. Each star's photometric and 
radial velocity membership flag, reproduced from Tables~\ref{table-identifier} 
\&~\ref{table-Prot}, are cited in columns 3~\&~4 respectively, whereas a 
flag confirming the detection of lithium 6708\r{A} in its spectrum (from 
Jones et al. 1997) is listed in column~5. Corresponding Jones et al. radial 
velocities are detailed in column~6, as well as their associated membership 
probabilities in column~7 (see below and Fig.~\ref{J97-RV-gauss}). V-band 
magnitudes of each star are listed in column~8 and finally, cluster 
membership probabilities based on the proper motions reported in JP96 
are reproduced in column~9.

In the first instance, we note that all stars in common between our gyrochronology sample 
and the Jones et al. sample have a significant lithium detection, indicative of stellar 
youth. These stars are therefore unlikely to be field star interlopers, which when combined 
with the photometric membership criterion, can safely be considered as {\em bona fide} cluster 
members. In the second, the radial velocity data from Jones et al. study do seem to cluster 
around -10km~s$^{-1}$, however, specific membership assignments are difficult to assess 
due to the associated scatter in the individual measurements. In order to understand 
whether the scatter in these radial velocities is due to a true dispersion in the data 
or is an admixture of single and binary member measurements, together with cluster 
non-members, we have constructed a histogram for all radial velocities published in the 
Jones et al. survey. This histogram, plotted in Fig.~\ref{J97-RV-gauss}, clearly 
shows a peak in the distribution representing the cluster's systemic heliocentric 
radial velocity. In order to investigate the dispersion about the central peak, we 
fit an unweighted Gaussian function to the velocity histogram, noting that the fit 
does not include background contamination due to binarity or cluster non-members, 
and assumes that the sample is complete. The Gaussian function is centred on $-12.49$ 
km~s$^{-1}$ with a 1$\sigma$ width of 2.32 km~s$^{-1}$, which we can exploit to determine 
a membership probability (reported in column~7 of Table~\ref{GYRO-vs-Jones97}) for each 
individual star based on its radial velocity and these Gaussian fit parameters. 

\subsection{Results:}

Analysis of the additional membership data reported in Table~\ref{GYRO-vs-Jones97} 
yields few surprises. Of all the Lowell periodic variables identified as 
photometric and/or RV members in Table~\ref{table-identifier}, only two 
of them, F3\_0172 \& F3\_0215, have a zero probability of being proper 
motion members according to the JP96 survey. Interestingly however, 
their V-magnitudes lie right at the point at which the JP96 proper 
motions begin to become less reliable due to Poisson-error limits on 
their photographic plate measurements, which brings their validity into 
question. Furthermore, we have noted in \S~\ref{GYRO-errors} that the 
gyrochronology sample may indeed suffer from contamination due to one 
or two cluster non-members masquerading as {\em bone fide} members, but we 
argue that their contribution to the period variance, and hence gyro 
errors, is small. In fact, F3\_0172 \& F3\_0215 contribute a combined 
period variance of 0.79 days$^{2}$, which is only a $1.8\%$ effect 
(see Table~\ref{Istars-table-Prot}). In spite of their low 
contribution to the period variance, we remain in the uncomfortable 
position of choosing whether to include these photometric and kinematic 
members of the cluster in our gyrochronology sample, or to exclude 
them on the basis of their proper motion membership probabilities. 
In order to further assess their cluster membership status, we 
have recently obtained high resolution optical spectra of these two 
stars using the Canada France Hawai`i Telescope [CFHT], the analysis 
of which we discuss below (see \S~\ref{cfhtdata}). 

Finally, all bar one of the Jones et al. (1997) stars are radial 
velocity members of the cluster, which correlates well with our own 
radial velocity data. The one Jones et al. star, F4\_0136, that is 
formally a cluster non-member based on a Gaussian probability fit 
to their radial velocity data plotted in Figure~\ref{J97-RV-gauss}, 
actually shows up as a single cluster star member in the long term 
synoptic velocity survey of Meibom et al. (in prep - see also 
\S~\ref{extantOBS}). Curiously, over the 9-epochs of observation 
covering 2.5-years, that Meibom et al. have for this star, its 
varies about {\em their} average radial velocity of -8.0 km~s$^{-1}$ 
by only 0.68 km~s$^{-1}$. Interestingly, a Gaussian fit to all radial 
velocities for the 70 single and binary M\,34 stars for which Meibom 
et al. have obtained results yields a cluster systemic velocity of 
$-7.59\pm1.02$ km~s$^{-1}$. Assuming the Jones et al. velocity is 
not in error, this star may be a binary member of the cluster, 
albeit with either a long-period orbit or a considerably eccentric 
one.

\subsection{Cluster Membership Status for stars F3\_0172 \& F3\_0215:}\label{cfhtdata}

Two stars exhibiting variability in our differential photometric 
survey of the M\,34 cluster, for which we derive periods, namely 
F3\_0172 \& F3\_0215, present us with somewhat of a conundrum. 
While these stars have photometric and radial velocity properties 
consistent with cluster membership, with relatively short photometric 
periods indicative of youth (compared to the Galactic field), and 
result in gyrochronology ages appropriate for a 200~Myr group of stars, 
they possess proper motion vectors incongruent with the remainder 
of the cluster. In an attempt to establish or refute genuine cluster 
membership for these two objects, we have recently observed them at 
high resolution using the fibre-fed, bench-mounted ESPaDOns \'{e}chelle 
spectrograph, located in a Coud\'{e}-like instrument chamber in the 
CFHT Observatory. The primary goal of these observations is to detect 
lithium at 6708\r{A}, and measure its equivalent width in these stars, 
thereby confirming their relative youth and increasing their probability 
of being {\em bona fide} cluster members.

During the evening of 27 January 2010, high resolution (R$\simeq60,000$) 
spectra were acquired for stars F3\_0172 \& F3\_0215 using the fibre-fed 
ESPaDOns spectrograph, in service with the 3.6-m Canada France Hawai`i 
Telescope located on top of Mauna Kea, Hawai`i, USA. The ESPaDOns 
spectrograph consists of a 79 lines mm$^{-1}$ \'{e}chelle grating imaged 
onto a $2048\times4608$ EEV CCD detector, having 13.5 $\mu$m square 
pixels, with photon input delivered by separate 100 $\mu$m (1.6~arcsec) 
diameter sky and target fibres. This set-up yields a FWHM of cross 
correlated ThAr arc lines of 0.271\r{A} at 6700\r{A}, and a complete 
wavelength range of $3699\rightarrow10481$\r{A}. Using this set-up, 
F3\_0172 \& F3\_0215 were observed for exposures totalling 2000 \& 2140 
seconds respectively, resulting in spectra with S/N$\simeq20$ at 6700\r{A}.

For each of the targets, we exploit our CFHT ESPaDOns spectra in order 
to measure their heliocentric radial velocity and their \lii 6708\r{A} 
EW. The removal of CCD instrumental effects, as well as the 
extraction of wavelength-calibrated spectra, have been achieved by 
two independent data reduction methodologies. In the first, we 
performed the bias-subtraction, flat-fielding, spectral order tracing, 
optimal extraction and wavelength calibration using standard IRAF 
procedures. In the second, we used the direct output from Libre-ESpRIT 
(Donati et al. 1997), the dedicated pipeline software for the ESPaDOns 
spectrograph. A cross-match of radial velocities and EWs for each 
target from the two reduction methods reassuringly yields consistent 
results to within $\pm 0.1$ km~s$^{-1}$ and $\pm$3m\r{A} respectively.

Heliocentric radial velocities were derived, relative to the well-exposed 
standard star HD 32963, in the spectral region of the Mg triplet lines 
($5104\rightarrow5176$\r{A}). Cross correlation of HD 32963 with other 
radial velocity standard stars observed during the CFHT programme shows 
that the zero-point in placing our velocities onto the standard system 
is $\simeq$0.2 km~s$^{-1}$. The relatively low S/N of our target spectra 
results in radial velocity precision errors of $\simeq$0.5 km~s$^{-1}$.

In order to measure equivalent widths for the \lii 6708\r{A} line in 
our targets, each spectrum between  $6586\rightarrow6711$\r{A}, was 
normalized using continuum fitting after spectra extraction. Each EW 
was calculated using both the direct integration and the Gaussian 
fitting methods, whose values were within a few percent of each 
other. In the lithium region, \lii EWs include contributions from 
the small Fe {\sc i}+CN line at 6707.44 \r{A}, leading to measured 
\lii EWs which are representative of a slightly ($10-20$ m\r{A}) 
over-estimated photospheric Li presence. Soderblom et al. (1993b) report 
that this Fe line blend has an EW = [20(B-V)$_{0}$ - 3]m\r{A}, determined 
through an empirical relationship for main sequence, solar-type stars. For 
each target star, we removed the \fei line contribution before transforming 
\lii EWs into abundances, N(Li) - on a scale where log N(H)=12, using the 
effective temperature-colour (B$-$V) relation from Soderblom et al (1993c), 
and the curves of growth presented in Soderblom et al. (1993b). Data products, 
radial velocity and lithium measurements, from the CFHT spectroscopic observations 
of F3\_0172 \& F3\_0215 are presented in Table~\ref{CFHT-table}.

CFHT radial velocities of F3\_0172 \& F3\_0215 are consistent with cluster 
membership of M\,34 irrespective of whether we compare their individual values 
to the Jones et al. (1997) sample or to the Meibom et al. (in prep) one. For 
the Jones et al. sample, their kinematic membership probabilities are non-zero 
although they are quite low at $08\%$ \& $03\%$ (for F3\_0172 \& F3\_0215 
respectively). Their membership probabilities are far more convincing when 
compared to the Meibom et al. sample (with respective values of 
$79\%$ \& $>$99\%).

\begin{figure*}
\includegraphics[angle=-90,width=16.4cm,clip]{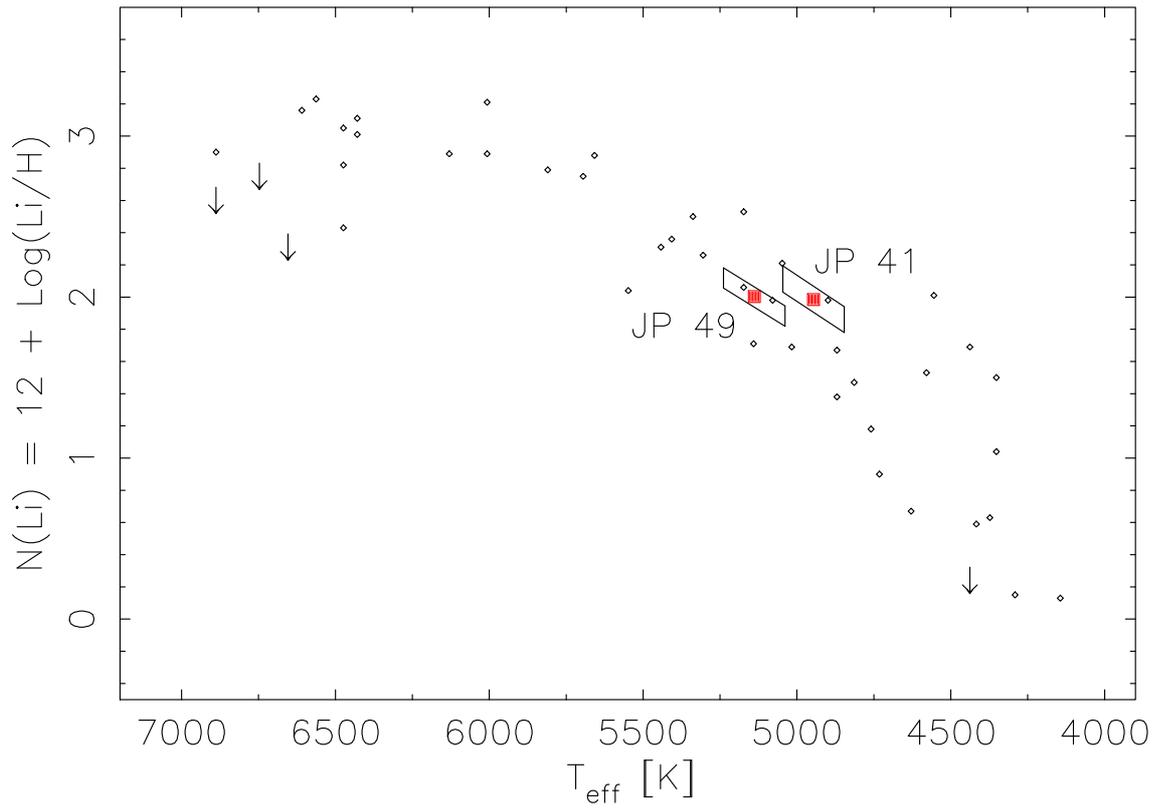}
\caption{Logarithmic lithium abundances N(Li) = 12 + log(Li/H) 
versus effective temperature are plotted for stars in the M\,34 cluster 
(data taken from Jones et al. 1997). Red filled squares represent 
abundances for F3\_0172 \& F3\_0215 (JP 49 \& JP 41), determined using 
data reported in Table~\ref{CFHT-table}, with error parallelograms 
based on temperature errors of 100~K and $15\%$ equivalent width 
uncertainties.}
\label{M34-lithium}
\end{figure*}

The lithium content of both F3\_0172 \& F3\_0215 is substantial indicating that 
these stars are relatively young compared to the general Galactic field, whose 
solar-type stellar content is typically old and has had sufficient time to have 
proton-burned considerable fractions of its natal lithium. In order for these 
stars to be judged as likely members of the M\,34 cluster, not only must they 
contain lithium in their atmospheres, but it must quantitatively fit into 
the mass-dependent lithium abundance distribution for the cluster. In 
Figure~\ref{M34-lithium}, we plot the lithium abundances for F3\_0172 \& 
F3\_0215 that we have determined from our CFHT spectra in concert with the 
extant mass-dependent lithium distribution for M\,34 stars (data taken from 
Jones et al. 1997, who employed identical (B-V)$_{o}$-Temperature, \fei 
line correction and lithium curves of growth as we have used). It is clear 
that both F3\_0172 \& F3\_0215 have lithium abundances consistent with the 
mass-dependent lithium distribution of the M\,34 cluster, and must be 
considered lithium abundance members.

JP96 proper motion vectors for F3\_0172 \& F3\_0215 indicate cluster 
non-membership. However their V-magnitudes render JP96 2-d kinematic 
membership probabilities questionable, because at or about this 
magnitude, JP96 proper motion accuracy and precisions begin to have 
strong dependencies on the Poissonian errors of their centroiding 
measurements. In consideration that these two stars are both photometric 
and kinematic members of the cluster, have photometric periods consistent 
with the remainder of the cluster's distribution, and have measured lithium 
abundances which lie right along the trend of the mass-dependent lithium 
distribution for the cluster, they are probable M\,34 cluster members 
and we retain them in our gyrochronology sample.

\clearpage

\begin{table*}
\caption{Cluster membership assessments for Jones \& Prosser (1996) stars in our period sample.}
\label{GYRO-vs-Jones97}
\begin{tabular}{rllcclccc}        
\hline\hline                 
Internal$^{a}$ & JP96$^{b}$ & Phot$^{a}$ & RV$^{a}$ & Li Mem$^{c}$ & RV J97$^{c}$ & RV Mem$^{d}$ & V$^{a}$ & PM Prob$^{e}$ \\
Identifier     &            & Mem        & Mem      & J97          & [km/s]       & J97          & [mag]   & JP96    \\
\hline
F3\_0172       & JP  49     & Y          &   Ys     &  ...         &  ...         & ...          & 14.584  & 00\%   \\
F3\_0176       & JP 167     & Y          &   Ys     &  ...         &  ...         & ...          & 14.659  & 44\%   \\
F3\_0215       & JP  41     & Y          &   Ys     &  ...         &  ...         & ...          & 14.918  & 00\%   \\
F3\_0258       & JP 289     & Y          &   Ys     &  Y           &  -10.8       & Y (47\%)     & 15.236  & 76\%   \\
F3\_0306       & JP 265     & Y          &   Yb     &  Y           &  -17         & Y ( 5\%)     & 15.503  & 53\%   \\
F3\_0320       & JP 172     & Y          &   ...    &  Y           &   -9.8       & Y (25\%)     & 15.664  & 03\%   \\
F4\_0136       & JP 213     & Y          &   Ys     &  Y           &   -5.7       & N ( 0\%)     & 13.151  & 94\%   \\
F4\_0169       & JP  50     & Y          &   Ys     &  ...         &  ...         & ...          & 13.510  & 04\%   \\
F4\_0194       & JP 133     & Y          &   Ys     &  Y           &  -11.8       & Y (77\%)     & 13.662  & 96\%   \\
F4\_0234       & JP 148     & Y          &   Ys     &  ...         &  ...         & ...          & 13.959  & 05\%   \\
F4\_0303       & JP 227     & Y          &   Ys     &  Y           &  -15.4       & Y (21\%)     & 14.366  & 92\%   \\
F4\_0317       & JP 224     & Y          &   Ys     &  Y           &   -9.4       & Y (18\%)     & 14.410  & 91\%   \\
F4\_0327       & JP 199     & Y          &   Ys     &  Y           &   -9.0       & Y (13\%)     & 14.483  & 85\%   \\
F4\_0335       & JP 158     & Y          &   Yb     &  Y           &   -9.9       & Y (26\%)     & 14.546  & 86\%   \\
F4\_0450       & JP 113     & Y          &   Ys     &  Y           &  -13.5       & Y (67\%)     & 14.960  & 84\%   \\
F4\_0667       & JP  52     & Y          &   Yb     &  ...         &  ...         & ...          & 15.623  & 26\%   \\
F4\_0695       & JP 197     & Y          &   Ys     &  ...         &  ...         & ...          & 15.643  & 05\%   \\
F4\_0730       & JP  18     & Y          &   ...    &  ...         &  ...         & ...          & 15.741  & 36\%   \\
F4\_0803       & JP 229     & Y          &   Ys     &  Y           &   -9.9       & Y (26\%)     & 16.025  & 46\%   \\
F3\_0121       & JP 212     & N          &   ...    &  ...         &  ...         & ...          & 13.741  & 00\%   \\
F3\_0163       & JP 258     & Y          &   N      &  ...         &  ...         & ...          & 14.440  & 00\%   \\
F3\_0165       & JP 244     & Y          &   N      &  ...         &  ...         & ...          & 14.419  & 00\%   \\
F3\_0388       & JP 174     & N          &   ...    &  ...         &  ...         & ...          & 15.952  & 00\%   \\
F3\_0413       & JP 218     & Y          &   N      &  ...         &  ...         & ...          & 16.242  & 17\%   \\
F4\_0167       & JP 223     & N          &   ...    &  ...         &  ...         & ...          & 13.502  & 00\%   \\
F4\_0404       & JP 131     & Y          &   N      &  ...         &  ...         & ...          & 14.730  & 00\%   \\
F4\_0515       & JP 129     & N          &   ...    &  ...         &  ...         & ...          & 15.111  & 00\%   \\
\hline
\end{tabular} \\

\noindent $a -$ Data taken from Tables~\ref{table-identifier} \& ~\ref{table-Prot}.\\
\noindent $b -$ {\em JP} identifier from Jones \& Prosser (1996).\\
\noindent $c -$ Cluster membership based on radial velocities and lithium 6708\r{A} 
equivalent widths, taken from Jones et al. (1997) [J97].\\
\noindent $d -$ Radial velocity membership assessments are based on a Gaussian fit to the 
entire Jones et al. (1997) radial velocity sample (see also Fig.~\ref{J97-RV-gauss}).\\
\noindent $e -$ Proper motion membership probabilities are taken from JP96.\\
\end{table*}

\clearpage

\begin{table*}
\caption{CFHT spectroscopic data products for M\,34 candidate members F3\_0172 \& F3\_0215.}
\label{CFHT-table}
\begin{tabular}{rllccccccc}        
\hline\hline                 
Internal$^{a}$ & JP96$^{b}$ & (B-V)$_{o}^{a}$ & Temp$^{c}$ & HJD$^{d}$    &  RV           & \lii + \fei$^{e}$ & \fei$^{f}$  & \lii 6708\r{A}$^{g}$ & N(Li)$^{h}$ \\
Identifier     &            &                 & [K]        & [(+2450000)] & [km~s$^{-1}$] &  EW [m\r{A}]      & EW [m\r{A}] & EW [m\r{A}]          &         \\ \hline
F3\_0172       & JP  49     &  0.811          & 5139       & 5223.713     & -8.4          &  115              & 13          & 102                  & 2.003       \\ 
F3\_0215       & JP  41     &  0.874          & 4946       & 5223.712     & -7.4          &  153              & 14          & 139                  & 1.985       \\
\hline
\end{tabular} \\

\noindent $a -$ Data taken from Tables~\ref{table-identifier} \& ~\ref{table-Prot}.\\
\noindent $b -$ {\em JP} identifier from Jones \& Prosser (1996).\\
\noindent $c -$ Temperature determined from (B-V)$_{o}$ relationship in Soberblom et al. (1993c - equ.~3) \\
\noindent $d -$ HJD - heliocentric julian date \\
\noindent $e -$ Equivalent width measurement, by Gaussian fitting, of the blended \lii and \fei 6708\r{A} lines. \\
\noindent $f -$ \fei 6707.441\r{A} line contribution, determined by an empirical (B-V) colour fit 
(see Soderblom et al. 1993b). \\
\noindent $g -$ \lii 6708\r{A} minus the \fei 6707.441\r{A} contribution.\\ 
\noindent $h -$ Lithium equivalent widths are transformed into abundance [N(Li)] - on a scale where log N(H)=12, 
using a curve of growth in Soderblom et al. (1993b). \\
\end{table*}

\clearpage

\clearpage

\clearpage

\section{A Period vs Equatorial Velocity Analysis}\label{Vsini-comp-period}

\begin{figure*}
\centering
\includegraphics[angle=-90,width=16.4cm,clip]{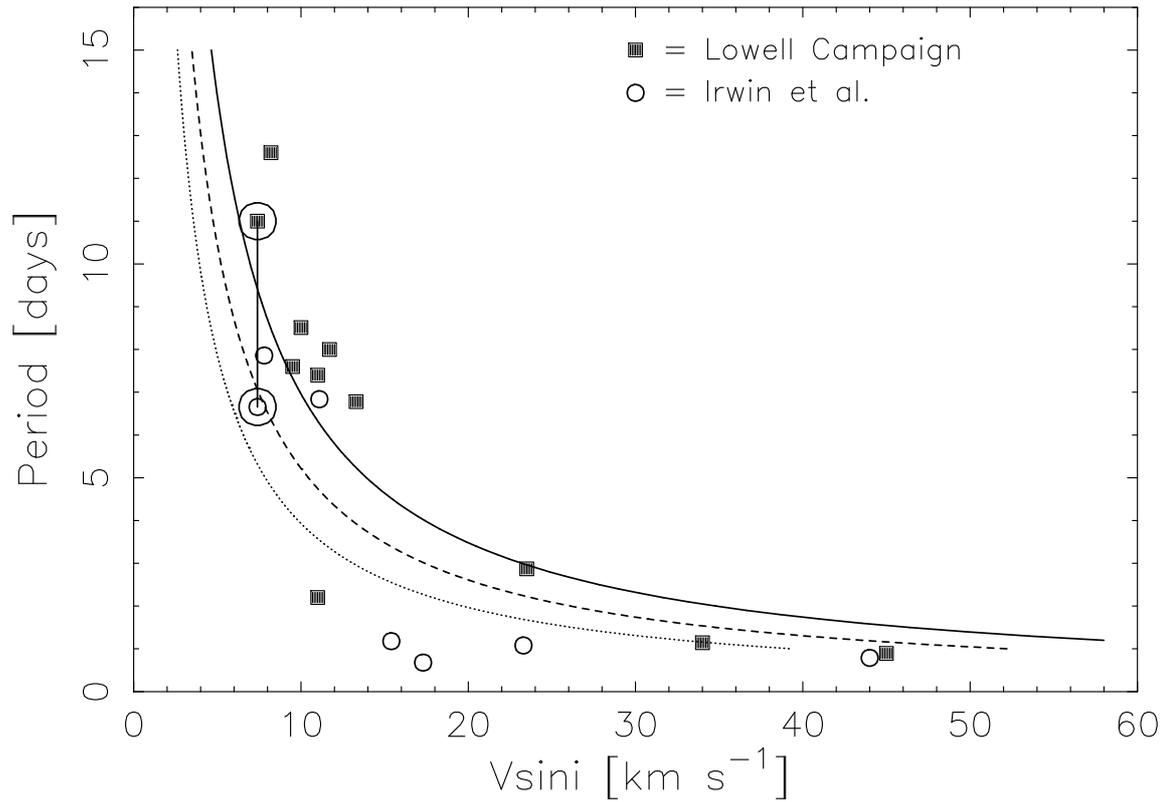}
\caption{Projected equatorial velocities ($v \sin i$) are plotted for those 
M\,34 stars having corresponding photometric periods derived during the 
course of the Lowell campaign. $V \sin i$ data are taken from Jones et al. 
(1997). For those cases where there is no Lowell campaign period available, 
similar data from the Irwin et al. (2006) sample are presented. 
One exception is shown for the star F3\_0258, linked by a straight 
solid line, whose Lowell campaign and Irwin et al. periods are seriously 
discrepant ($11.0 \pm 1.0$  and 6.655 days respectively). Three loci 
are depicted representing equal period-vs-\rot relationships 
(assuming sin$i$ = $\pi/4$) for G0, K0 and M0 dwarfs (solid, 
dashed and dotted lines respectively). In order of descending mass, 
stellar radii of R/R$_{\odot}$ = 1.08, 0.81 \& 0.61 were employed 
(Gray 1992).}
\label{VSINI}
\end{figure*}

A comparison of photometric period and projected 
equatorial rotation rate for M\,34 stars is plotted 
in Fig.,~\ref{VSINI}. For the most part, photometric 
periods are determined from the Lowell campaign 
described in this manuscript, except for a few cases 
where Irwin et al. periods are employed where Lowell 
ones do not exist. Spectroscopic velocities for stars 
in common to the Lowell campaign and Irwin  et al. 
survey are obtained from Jones et al. (1997). Three 
loci are also shown in the figure, representing 
equality between period and \rot data for G0, K0 \& 
M0 dwarfs, where $\sin i = \pi/4$ is assumed for each. 

There are three notable characteristics to the plot. 
First, the clustering of data-points for stars 
with periods $\simeq 6.5\rightarrow9$ days and \rot 
$\simeq 10-14$ km~s$^{-1}$ lie to the right of the 
G0 dwarf locus. Given that the M\,34 stars having 
photometric periods are late-F to early M-dwarfs, 
these data-points are most likely indicative of 
those stars whose $\sin i$ values are $> \pi/4$. 
Second, there are four stars with periods $<3$ days 
and $10<$\rot$<25$ km~s$^{-1}$, whose period,\rot 
data place them considerably below even the 
M0 dwarf locus in the diagram. Assuming that 
these stars are {\em bona fide} cluster members, 
whose period and \rot values are correctly 
determined, these stars must be high inclination 
systems ($\sin i < \pi/4$; \ie becoming more pole-on), 
whose true equatorial velocities are considerably 
higher. 

Finally, there is one M\,34 star (F3$\_0258$) whose 
period determinations, from our Lowell data and by 
the Irwin et al. study, are seriously discrepant 
(see also Fig.,~\ref{IRWINcompFIG}). The periodicity 
for this star was detected in three filter/exposure 
time observations during our Lowell campaign. It 
is interesting to note that the Irwin et al. period 
is almost half that of our Lowell one, and their 
lower value could be due to phase aliasing, power 
leakage in the power spectrum as a consequence of 
their shorter observing window or multiple spot groups 
on the surface on the star during its observation. 
In any case, either both period determinations are 
incorrect, or either value from the Lowell campaign 
or the Irwin et al. study is in error. 

If we assume that one of the periods for this star 
is correct, we can make some predictive statements 
as to its equatorial velocity. With an intrinsic 
B-V colour of 0.95, its spectral type on the main 
sequence would be K2 or K3, placing it close to 
the central locus of the three plotted in 
Fig.,~\ref{VSINI} (assuming the star is inclined 
$45^{o}$  to the line-of-sight). This scenario is 
consistent with its Irwin et al. period of 6.655 
days. Conversely, if the Lowell period of 11.0 days 
is correct for this object, and it is a single 
member of the cluster lying on the main sequence, 
its inclination angle must be higher than $45^{o}$ 
($\sin i > \pi/4$), whose appearance is more face-on 
to the line-of-sight. Hopefully, more extensive 
photometric monitoring of this star will reveal its 
true nature.

\clearpage

\section{Lowell Light-curves for Photometrically Variable Stars in 
the Field of M\,34}\label{lightcurves}

\begin{figure*}
\includegraphics[angle=-90, scale=0.65]{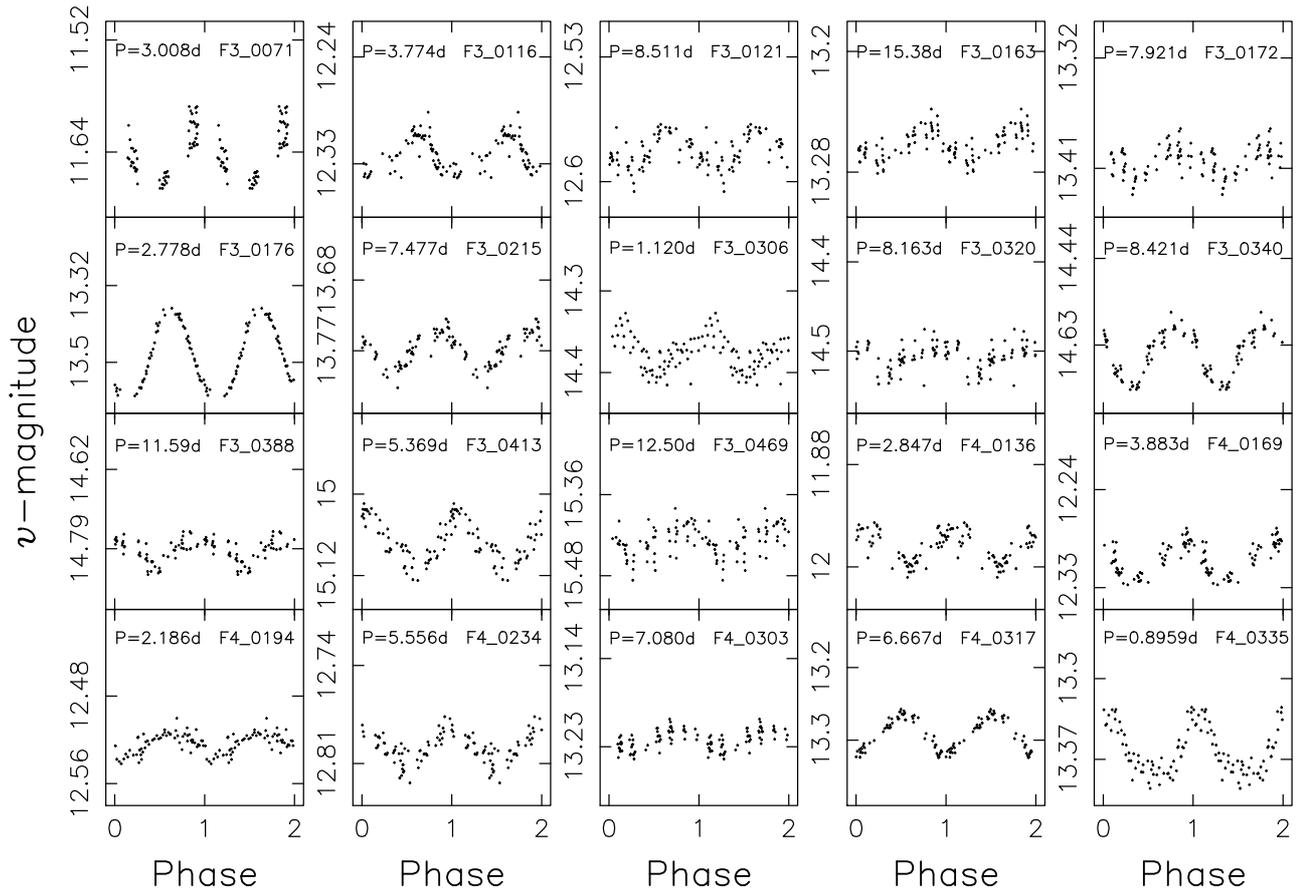}
\caption{Period-phased photometric lightcurves for M\,34 variables, derived 
from short-V observations.} 
\label{PERIOD-sV1to20}
\end{figure*}

\begin{figure*}
\includegraphics[angle=-90, scale=0.65]{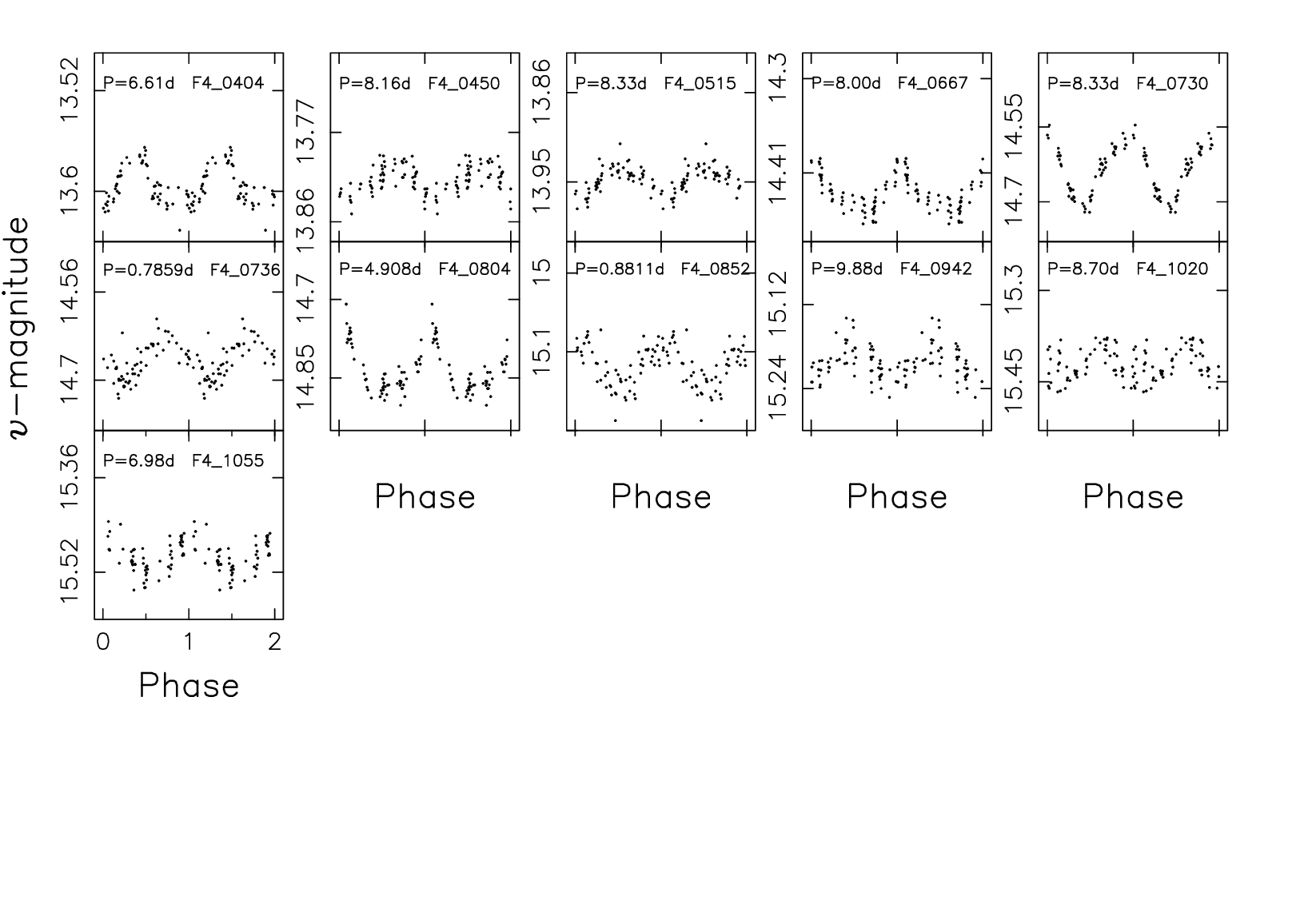}
{\bf Fig.~\ref{PERIOD-sV1to20}} Cont/d....
\label{PERIOD-sV21to31}
\end{figure*}


\begin{figure*}
\includegraphics[angle=-90, scale=0.65]{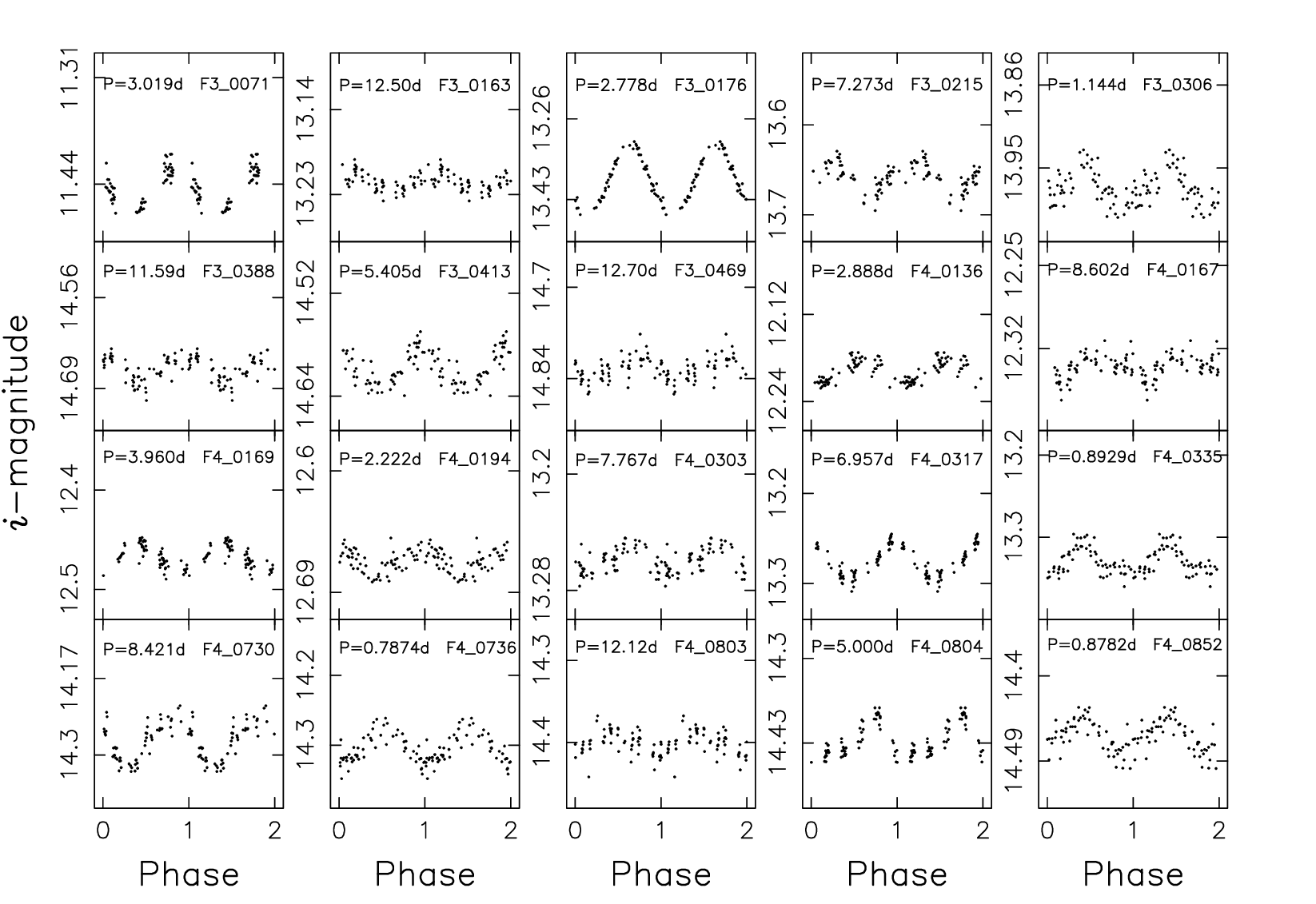}
\caption{Period-phased photometric lightcurves for M\,34 variables, 
derived from short-I observations.} 
\label{PERIOD-sI1to20}
\end{figure*}

\begin{figure*}
\includegraphics[angle=-90, scale=0.65]{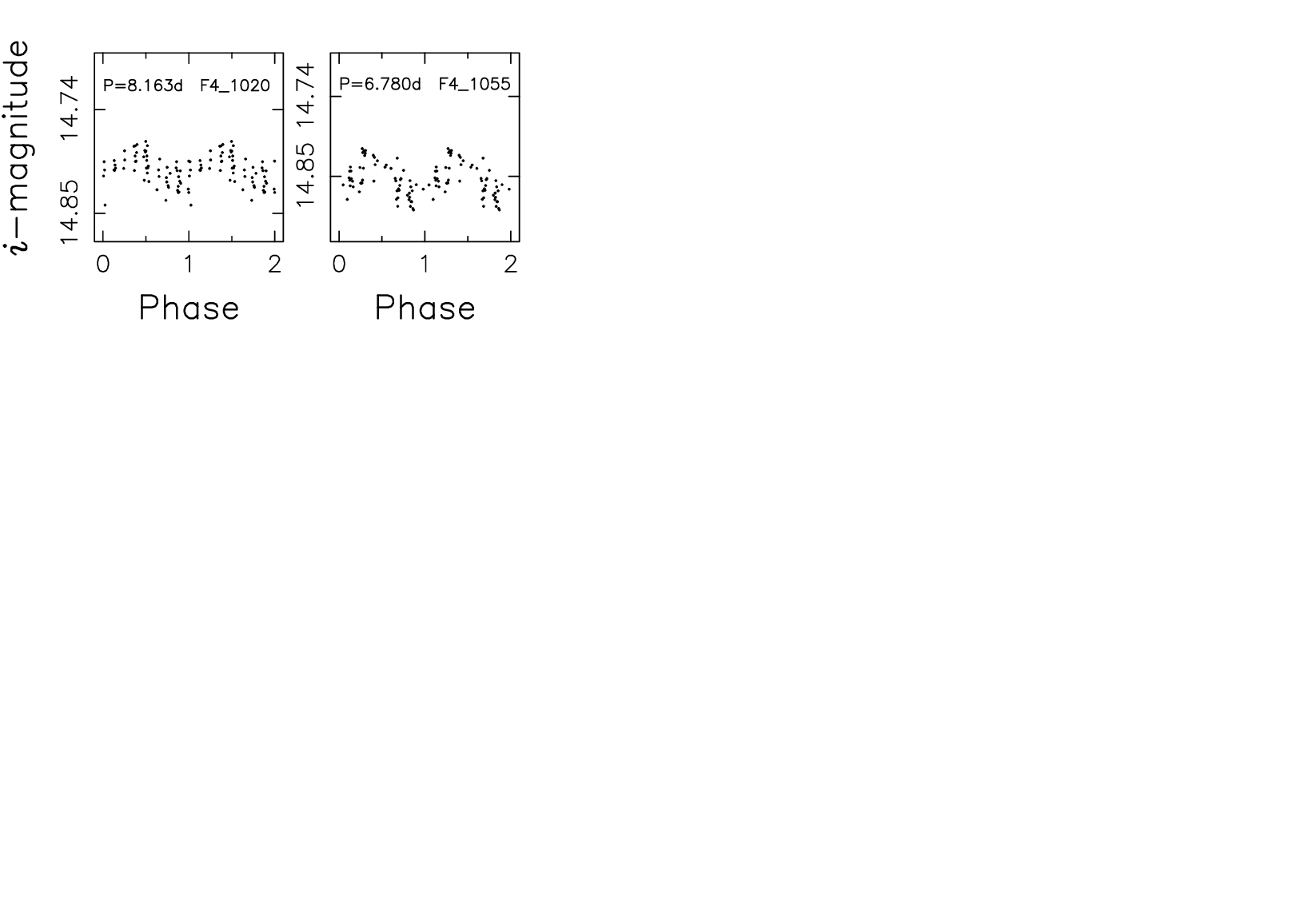}
{\bf Fig.~\ref{PERIOD-sI1to20}} Cont/d....
\label{PERIOD-sI21to22}
\end{figure*}


\begin{figure*}
\includegraphics[angle=-90, scale=0.65]{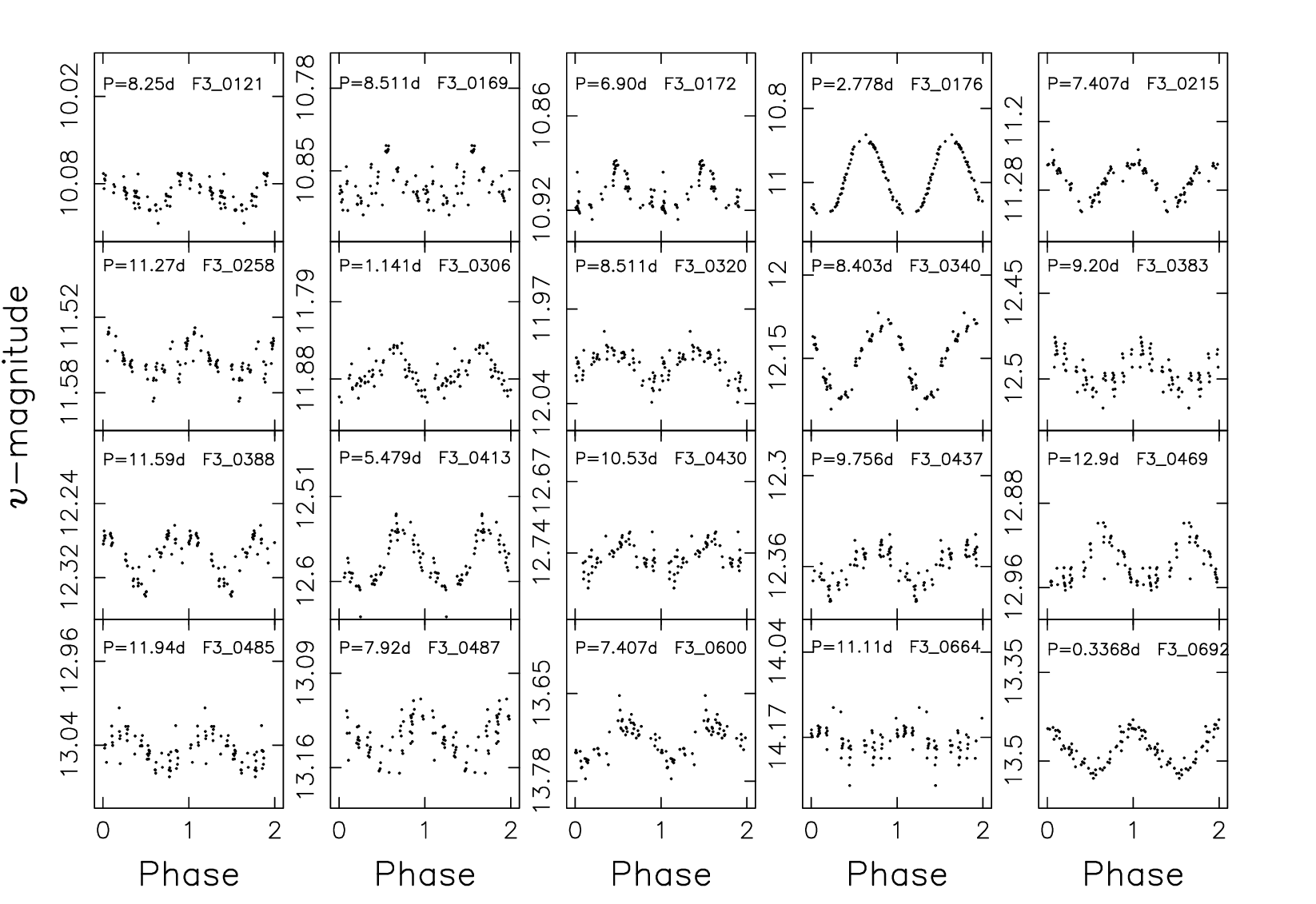}
\caption{Period-phased photometric lightcurves for M\,34 variables, 
derived from long-V observations.} 
\label{PERIOD-lV1to20}
\end{figure*}

\begin{figure*}
\includegraphics[angle=-90, scale=0.65]{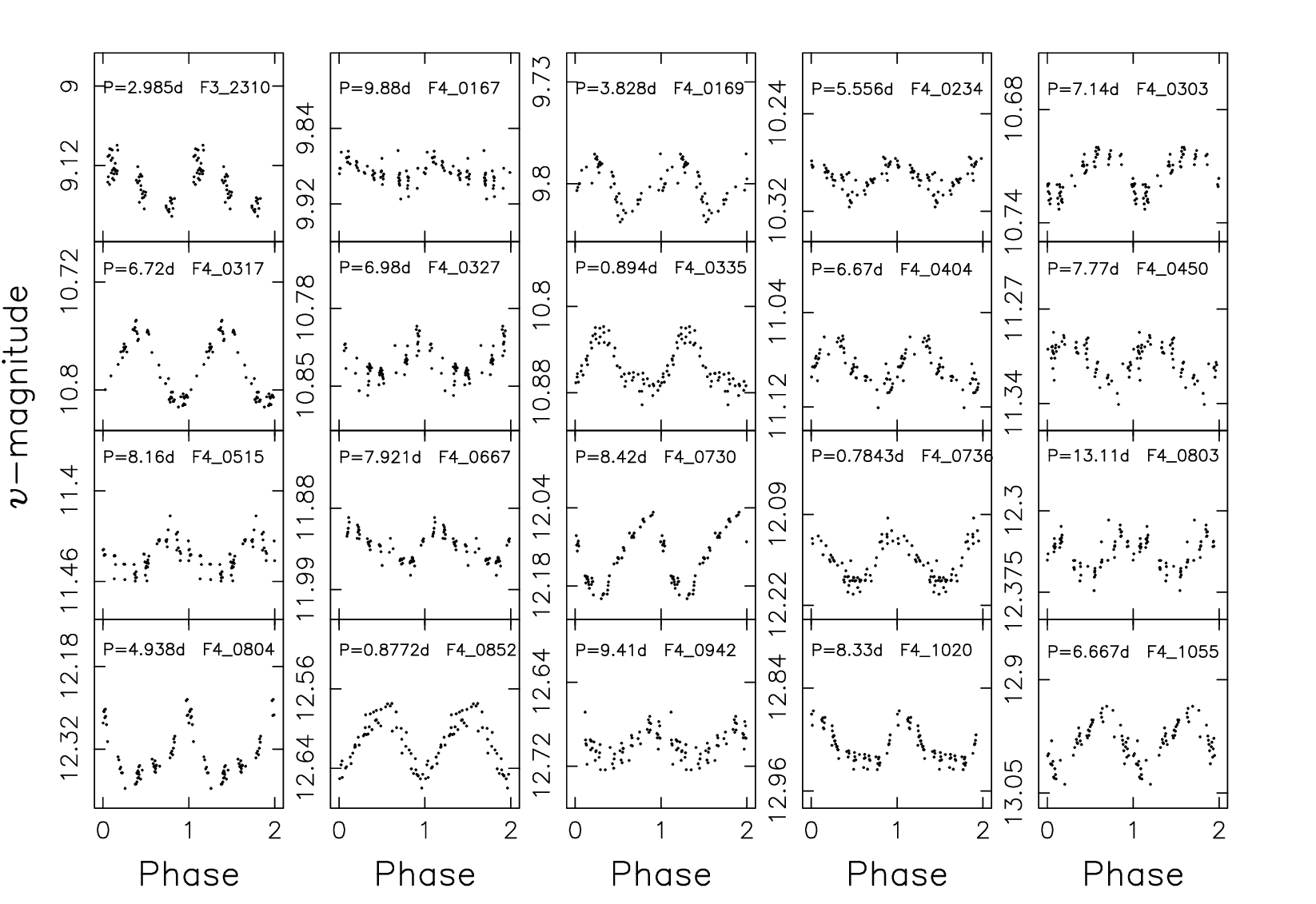}
{\bf Fig.~\ref{PERIOD-lV1to20}} Cont/d....
\label{PERIOD-lV21to40}
\end{figure*}

\begin{figure*}
\includegraphics[angle=-90, scale=0.65]{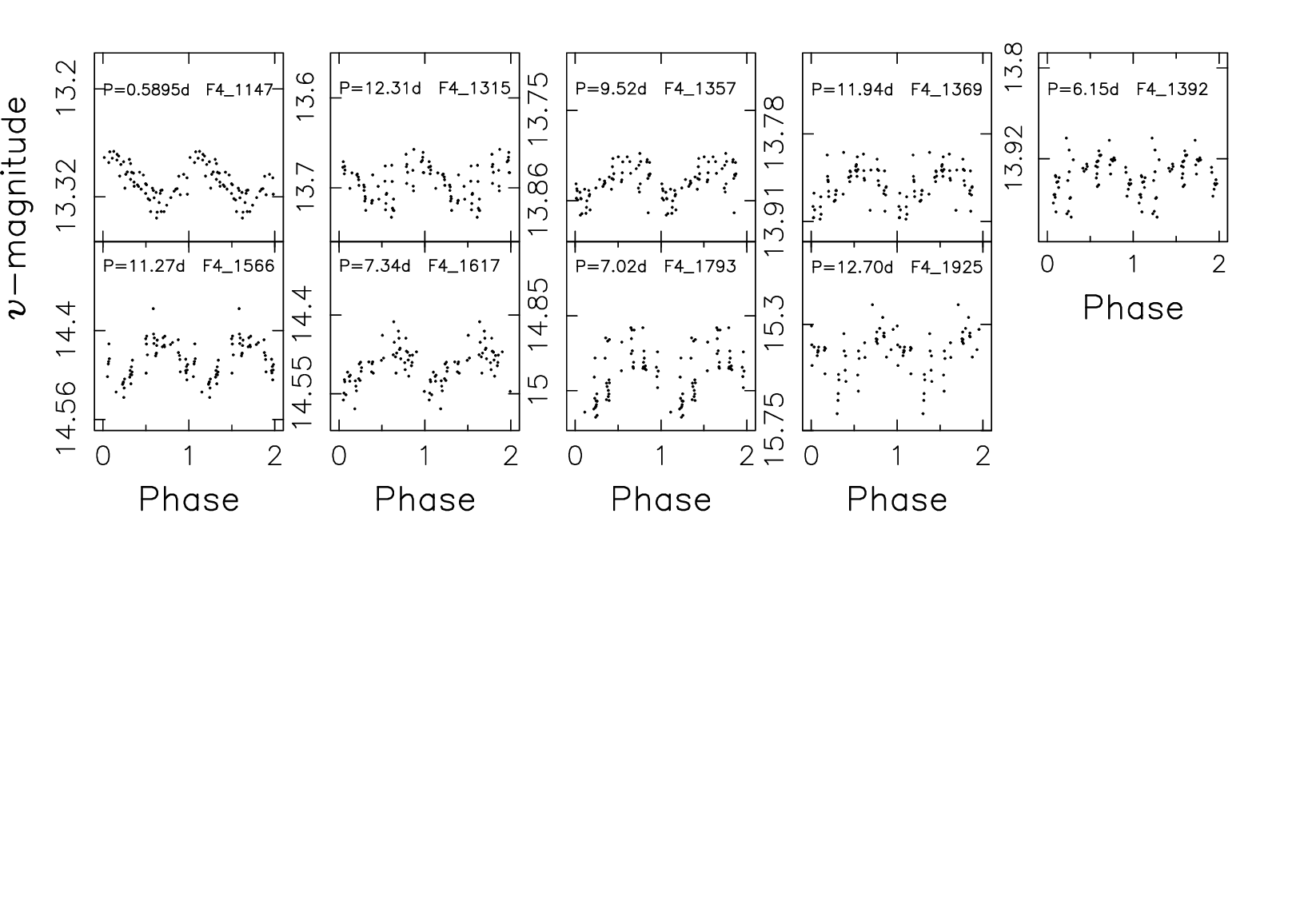}
{\bf Fig.~\ref{PERIOD-lV1to20}} Cont/d....
\label{PERIOD-lV41to49}
\end{figure*}

\begin{figure*}
\includegraphics[angle=-90, scale=0.65]{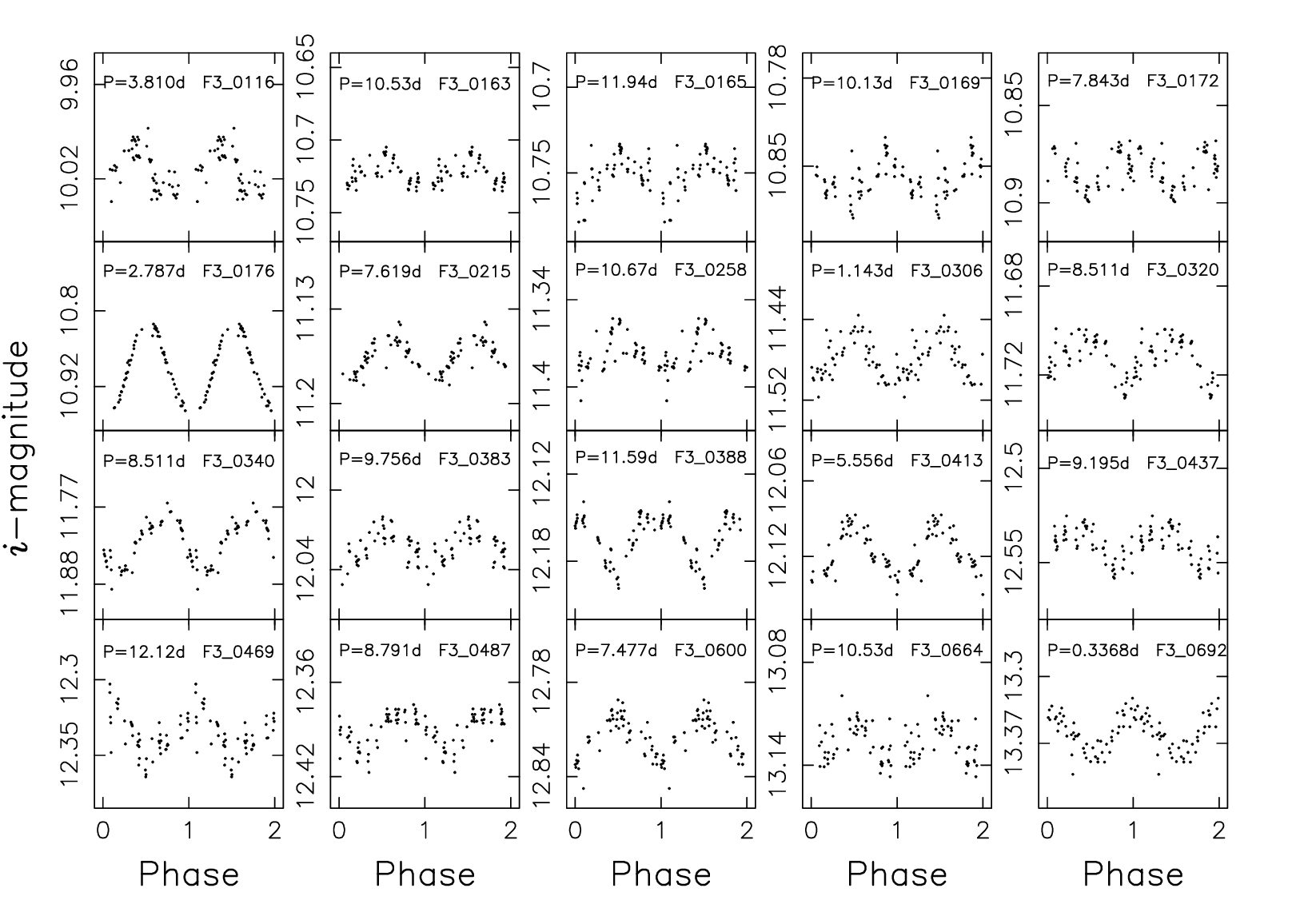}
\caption{Period-phased photometric lightcurves for M\,34 
variables, derived from long-I observations.} 
\label{PERIOD-lI1to20}
\end{figure*}

\begin{figure*}
\includegraphics[angle=-90, scale=0.65]{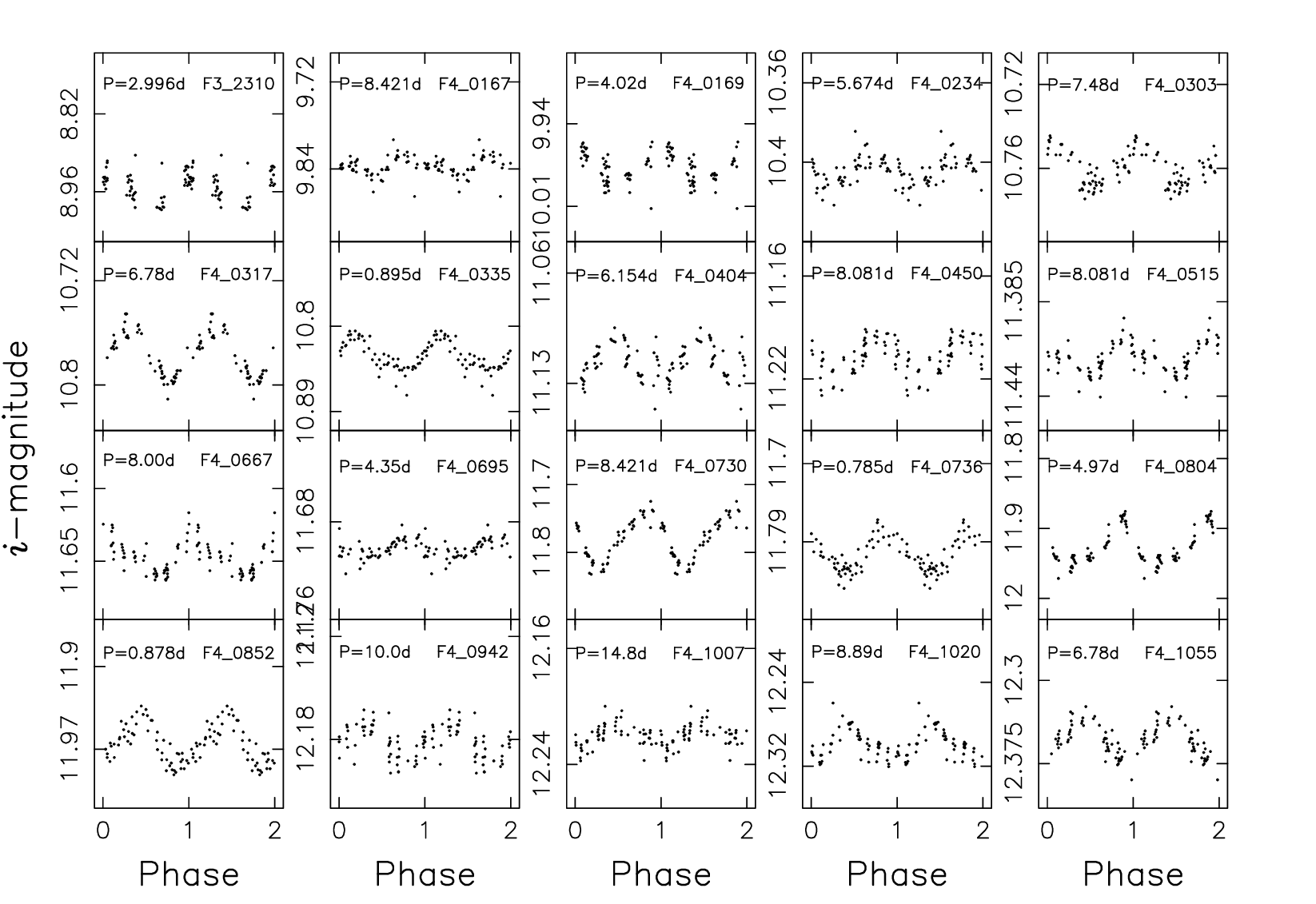}
{\bf Fig.~\ref{PERIOD-lI1to20}} Cont/d....
\label{PERIOD-lI21to40}
\end{figure*}

\begin{figure*}
\includegraphics[angle=-90, scale=0.65]{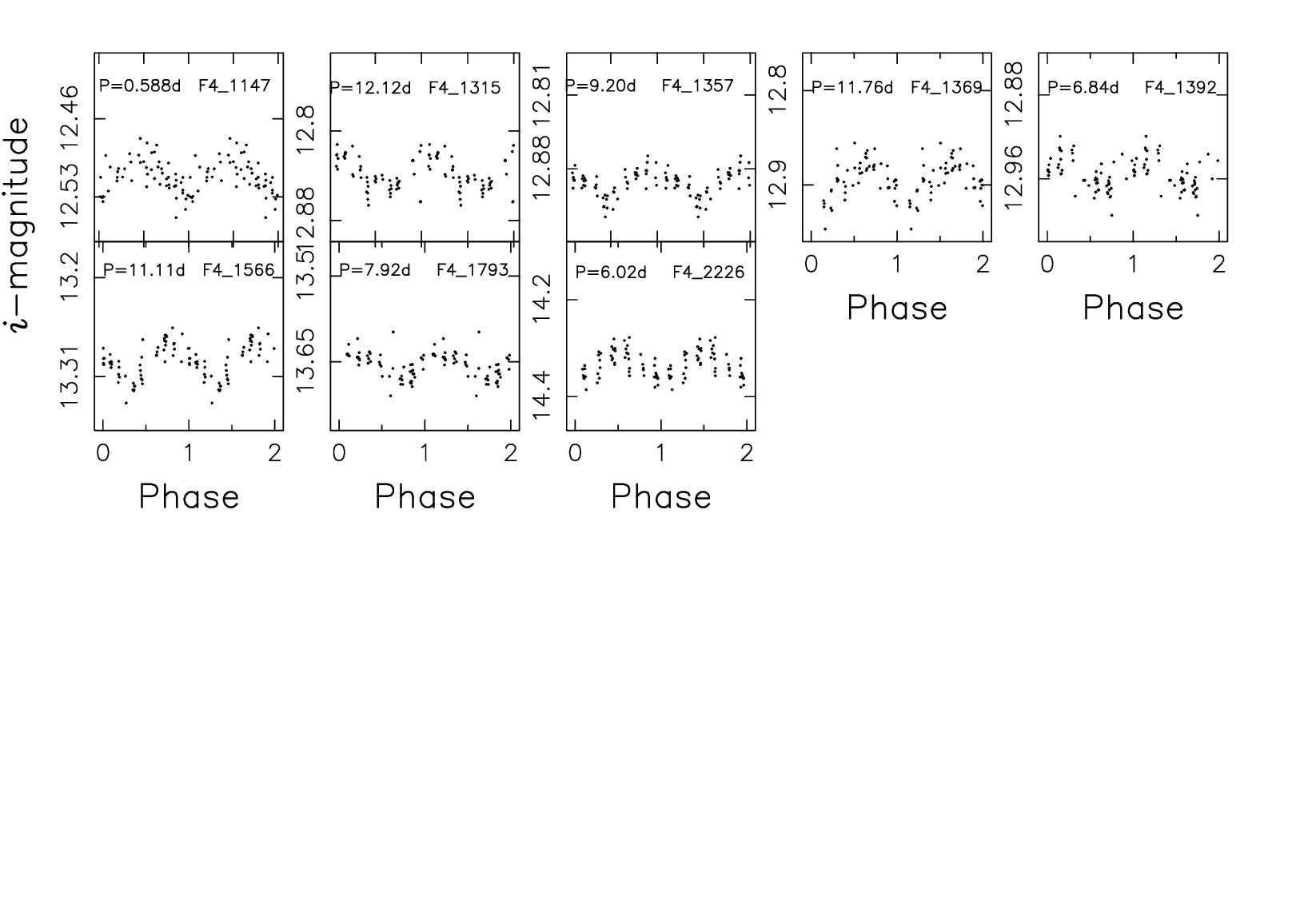}
{\bf Fig.~\ref{PERIOD-lI1to20}} Cont/d....
\label{PERIOD-lI41to48}
\end{figure*}

\end{appendix}

\end{document}